%% file: FrontiersMCARXIV.tex
\documentclass[utf8]{WKFPHY} 
\setcitestyle{square} 
\usepackage{url,hyperref,lineno,microtype,subcaption}
\usepackage[onehalfspacing]{setspace}
\usepackage[all,cmtip]{xy}
\usepackage{xspace}
\usepackage{tikz-cd}
\input{latexcommands_2021a.tex}

\newcommand{\FACTOR}[2]{\glb #1, #2 \grb}

\newcommand{\Conf}{\ensuremath{c}}

\newcommand{\NewConf}{\ensuremath{c'}}
\newcommand{\pMet}{\PCAL^{\text{Met} }}
\newcommand{\pFact}{\PCAL^{\text{Fact} }}

\newcommand{\RR}{\mathbb{R}}

\newcommand{\Lfree}{L_\text{free}}
\newcommand{\tvd}[1]{|| #1 ||_{\text{TV}}}
\newcommand{\tmix}[1][]{t_{\text{mix}}}
\newcommand{\tcorr}[1][]{t_{\text{corr}}}
\newcommand{\Npath}{n}
\newcommand{\pit}[1]{\pi^{\{#1\}}}
\newcommand{\Ptrans}{\Phat^{\text{trans}}}
\newcommand{\Pres}{\Phat^{\text{res}}}

\def\keyFont{\fontsize{8}{11}\helveticabold }
\def\firstAuthorLast{Krauth} 
\def\Authors{Werner Krauth$^{1,*}$}
\def\Address{$^{1}$Laboratoire de Physique de l’Ecole normale supérieure, ENS, 
Université PSL, CNRS, Sorbonne Université, Université Paris-Diderot, Sorbonne 
Paris Cité, Paris, France}
\def\corrAuthor{Corresponding Author}
\def\corrEmail{werner.krauth@ens.fr}

\begin{document}
\onecolumn
\firstpage{1}
\title[Event-chain Monte Carlo]{Event-chain Monte 
Carlo: foundations, applications, and prospects} 
\author[\firstAuthorLast ]{\Authors} 
\address{} 
\correspondance{LPENS, 24 rue Lhomond, 75005 Paris, 
France}
\extraAuth{}

\maketitle

\begin{abstract}
\section{}
This review treats the mathematical and algorithmic foundations of 
non-reversible Markov chains in the context of event-chain Monte Carlo (ECMC), 
a continuous-time lifted Markov chain that employs the factorized 
Metropolis algorithm. It analyzes a number of model applications, and then 
reviews the formulation as well as the performance of ECMC in key models in 
statistical physics. Finally, the review reports on an ongoing initiative to 
apply the method to the sampling problem in molecular simulation, that is, to 
real-world models of peptides, proteins, and polymers in aqueous solution.
\tiny
\keyFont{ \section{Keywords:} Event-driven simulations, Monte Carlo 
methods, non-reversible Markov chains, lifting, molecular simulation, molecular 
dynamics, Coulomb potential}
\end{abstract}

\section{Introduction}
\label{sec:Introduction}

Markov-chain Monte Carlo (MCMC) is an essential tool for the natural sciences. 
It 
is also the subject of a research discipline in mathematics. MCMC has from the 
beginning~\cite{Metropolis1953}, in 1953, focussed on reversible MCMC 
algorithms, those
that satisfy a detailed-balance condition. These algorithms are particularly 
powerful when the Monte Carlo moves (from one configuration to the next) can be 
customized for each configuration of a given system. Such \emph{a priori} 
choices allow for big moves to be accepted, and for sample space 
to be explored rapidly. Prominent 
examples for reversible methods with custom-built, large-scale, moves are 
path-integral Monte Carlo and the cluster algorithms for spin 
systems~\cite{SMAC}.

In many important problems, insightful \emph{a priori} choices for moves are 
yet unavailable. MCMC then often consists of a sequence of  unbiased local 
moves, such as tiny displacements of one out of $N$ particles, or flips of one 
spin out of many. Local reversible Monte Carlo schemes are easy to set up for 
these problems. The Metropolis or the heatbath (Gibbs-sampling) algorithms are 
popular choices. They generally compute acceptance probabilities from the 
changes in the total potential (the system's energy) and thus mimic the 
behavior of physical systems in 
thermodynamic equilibrium. However, such algorithms are often too slow to be 
useful. Examples are the hard-disk model~\cite{Alder1962,Bernard2011} where 
local reversible MCMC methods failed for several decades to obtain independent 
samples in large systems, as well as the vast field of molecular 
simulation~\cite{FrenkelSmitBook2001}. That field considers classical models of 
polymers and proteins, \emph{etc}., in aqueous solution. Local reversible MCMC 
algorithms were for a long time without alternatives in molecular simulation, 
but they remain of little use because of their extreme inefficiency.

Event-chain Monte Carlo (ECMC)~\cite{Bernard2009,Michel2014JCP} is a family of 
local, 
non-reversible MCMC algorithms that has been developed over the last decade. 
Its foundations, applications, and prospects are at the heart of the present 
review. At a difference with its local reversible counterparts (that are all 
essentially equivalent to  each other and to the physical overdamped 
equilibrium dynamics), different representatives of ECMC present a wide spread 
of behaviors for the same problem. Two issues explain this spread. First, any 
decision to accept or reject a given move is made on the basis of a 
consensus~\cite{Michel2014JCP} between statistically independent  factors, 
which each regroup parts of the total potential. Usually, there is a choice
between inequivalent factorizations. In the Metropolis algorithm, in contrast, 
all decisions to accept a move are made on the basis of changes in the total 
energy. Second, ECMC is fundamentally a non-reversible, \quot{lifted}, version 
of an underlying reversible algorithm. In a lifted Markov chain, some of the 
randomly sampled moves of the original (\quot{collapsed}) Markov chain are 
rearranged. Again, there are many options for rearranging
moves, and each of these can give rise to a specific dynamic behavior. The 
development of ECMC is still in its infancy. Its inherent variability may 
however bring to life the use of local MCMC  in the same way as reversible 
energy-based Monte Carlo has been empowered through the \emph{a priori} 
probabilities.

In the following, \sect{sec:MarkovChains} discusses the mathematical 
foundations underlying ECMC, namely global balance,  factorization, lifting, 
and thinning (making statistically correct decisions with minimal effort). 
\sect{sec:SingleParticles} reviews exact results for various Markov chains 
in the simplest setting of a path graph. \sect{sec:OneDimensionalN} studies 
one-dimensional $N$-body systems \sect{sec:StatMechgreater} provides an 
overview of findings on $N$-particle systems in higher than one dimension. 
\sect{sec:MolecularSimulation} reviews a recent proposal to apply 
ECMC to molecular simulations. \sect{sec:Prospects}, finally discusses 
prospects for event-chain Monte Carlo and other non-reversible 
Markov chains in the context of molecular simulation. \\

\section{Markov chains, factorization, lifting, and thinning}
\label{sec:MarkovChains}

The present section reviews fundamentals of ECMC. Non-reversibility 
modifies the positioning of MCMC  from an analogue of equilibrium physics 
towards the realization of a non-equilibrium system  with steady-state flows. 
\subsect{subsec:TransitionMatrix} discusses transition matrices, the objects 
that express MCMC algorithms in mathematical terms. The balance conditions for 
the steady state of Markov chains are reviewed, as well as the performance 
limits of MCMC in terms of basic characteristics of the sample space and 
the transition matrix. \subsect{subsec:Factorization} discusses factorization, 
the break-up of the interaction into terms that, via the factorized Metropolis 
algorithm,  make independent decisions on the acceptance or the rejection of a 
proposed move. \subsect{subsec:Lifting} reviews the current understanding 
of lifting, the mathematical concept that allows non-reversible Markov chains 
including ECMC to replace random moves by deterministic ones without modifying 
the stationary distribution and while remaining within the framework of
(memory-less) Markov chains. 
\subsect{subsec:Thinning} reviews thinning, the way used in ECMC to make 
complex 
decisions with minimum computations. \\

\subsection{Transition matrices, balance conditions}
\label{subsec:TransitionMatrix}

For a given sample space $\Omega$, a Markov chain consists in a 
time-independent 
transition matrix $P$. Its non-negative element  $P_{ij}$ gives the 
conditional probability for the configuration $i$ to move to configuration $j$ 
in one time step. The transition matrix $P$ is stochastic---all its rows $i$ 
sum 
up to $\sum_{j \in \Omega} P_{ij} = 1$. Commonly, the probability $P_{ij}$ is 
composed of two parts $P_{ij} = \ACAL_{ij} \PCAL_{ij}$, where $\ACAL$ is the 
\emph{a priori} probability to propose a move (mentioned in 
\sect{sec:Introduction}), and $\PCAL_{ij}$ the socalled \quot{filter} to accept 
the proposed move. Any term $P_{ii}$ in the above transition matrix is the 
probability to move from $i$ to $i$. If non-vanishing, it may result from the 
probability of all proposed moves from $i$ to other configurations $j \ne i$ 
(by 
the \emph{a priori} probability) to be rejected (by the filter $\PCAL$). In the 
lifted sample space $\Omegahat$ that will be introduced in 
\subsect{subsec:Lifting}, ECMC is without rejections. \\

\subsubsection{Irreducibility and convergence, basic properties of 
the transition matrix}
\label{subsec:TransitionIrreducibility}

A (finite) Markov chain
is irreducible if any configuration $j$ can be reached from any 
other configuration $i$ in a finite number of 
steps~\cite[Sect. 1.3]{Levin2008}.   As the matrix $P^t = (P^t)_{ij}$ gives the 
conditional probability to move 
from $i$ to $j$ in exactly $t$ time steps, an irreducible matrix  has 
$(P^t)_{ij} > 0$ $\forall i,j$, but the time $t$ may depend on $i$ and $j$. 

The transition matrix $P$ connects not only configurations but also probability 
distributions $\pit{t-1}$ and $\pit{t}$ at subsequent time steps $t-1$ and 
$t$. By extension, the  matrix $P^t$ connects the distribution $\pit{t}$ 
with the (user-provided) distribution of initial configurations $\pit{0}$ at 
time $t=0$:
\begin{equation}
\pit{t}_i = \sum_{j \in \Omega} \pit{t-1}_j P_{ji} \quad \Rightarrow 
\quad \pit{t}_i = 
\sum_{j \in \Omega} \pit{0}_j (P^t)_{ji}\quad \forall i \in \Omega.
\label{equ:TransitionMatrixDistribution}
\end{equation}
The initial distribution $\pit(0)$ can be concentrated on a single initial 
configuration. In that case, 
$\pit{0}$ is a discrete Kronecker $\delta$-function for a finite Markov chain, 
and a Dirac $\delta$-function otherwise.

An irreducible  Markov chain has a unique stationary distribution $\pi$ 
that satisfies
\begin{equation}
\pi_i = \sum_{j \in \Omega} \pi_j P_{ji}\quad \forall i \in \Omega.
\label{equ:GlobalBalance}
\end{equation}
\eq{equ:GlobalBalance} allows one to define the flow $\FCAL_{ji}$
from $j$ to $i$ as the stationary probability  $\pi_j$ to be at $j$ 
multiplied with the conditional probability to move from $j$ to $i$: 
\begin{equation}
\FCAL_{ji} \equiv \pi_j P_{ji} \quad \Rightarrow \quad 
\text{\eq{equ:GlobalBalance}} 
\Leftrightarrow 
\overbrace{\sum_{k \in \Omega} \FCAL_{ik}} ^{\text{flows exiting $i$}}
= 
\overbrace{\sum_{j \in \Omega} \FCAL_{ji}}^{\text{flows entering $i$}} \quad 
\forall i \in \Omega,
\label{equ:FlowExitEnter}
\end{equation}
where the left-hand side of \eq{equ:GlobalBalance}  was multiplied with the 
stochasticity condition $\sum_{k \in \Omega} P_{ik}=1$.

Although any irreducible transition  matrix  has a unique $\pi$, this  
distribution  is not guaranteed to be the limit $\pit{t}$ for 
$t \to \infty$ for all initial
distributions $\pit{0}$. But even in the absence of 
convergence, ergodicity follows from irreducibility alone, 
and ensemble averages 
$\sum_{i \in \Omega} \OCAL(i) \pi_i $ of an observable $\OCAL$ agree with the 
time averages $\lim_{t \to \infty} \frac{1}{t} \sum_{s=0}^{t-1} \OCAL(i_s)$ 
(see~\cite[Theorem 4.16]{Levin2008}).

Convergence towards $\pi$ of an irreducible Markov chain requires that it  is 
aperiodic,  that is, that the return times from a configuration $i$ back to 
itself $\SET{t\ge 1: (P^t)_{ii} > 0}$ are not all multiples of a period larger 
than one. For example, if the set of return times is $\SET{2,4,6,\dots}$, then 
the period is $2$, but if it is $\SET{1000, 1001, 1002, \dots}$, then it is 
$1$. 
These periods do not depend on the configuration $i$. For irreducible, 
aperiodic 
transition matrices, $P^t = (P^t)_{ij}$ is a positive matrix for some fixed 
$t$, 
and MCMC converges towards $\pi$ from any starting distribution $\pit{0}$. The 
existence and uniqueness of the stationary distribution $\pi$ follows from the 
irreducibility of the transition matrix, but if its value is imposed (for 
example to be the Boltzmann distribution or the diagonal density 
matrix~\cite{SMAC}), then the steady-state flow rate of  \eq{equ:GlobalBalance} 
becomes a necessary  condition for the transition matrix called \quot{global 
balance}. For ECMC, this global-balance condition of \eq{equ:GlobalBalance} 
must 
be checked for all members of the lifted sample space $\Omegahat$.

A reversible transition matrix is one that satisfies the detailed-balance 
condition 
\begin{equation} 
\FCAL_{ij} = 
\overbrace{\pi_i P_{ij}}^{\text{flow from $i$ to $j$}}= \overbrace{\pi_j 
P_{ji}}^{\text{flow from $j$ to $i$}}  = 
\FCAL_{ji} \
\forall i,j \in \Omega. 
\label{equ:DetailedBalance} 
\end{equation} 
Detailed balance implies global balance (\eq{equ:DetailedBalance} yields 
\eq{equ:GlobalBalance} by summing over $j$, considering that $\sum_{j \in 
\Omega} P_{ij} = 1$), and the flow into a configuration $i$ coming from a site 
$j$ goes right back to $j$.  In ECMC, the reality of the global-balance 
condition is quite the opposite, because the entering flow $\FCAL_{ji}$ is 
compensated by flows to other sites than $j$ ($\FCAL_{ji}> 0$ usually implies 
$\FCAL_{ij} = 0$). Checking the global balance condition is more complicated 
than checking detailed balance, as it requires monitoring all the 
configurations 
$j$ that contribute to the flow into $i$, in an augmented sample space.

For a reversible Markov chain, the matrix $A_{ij} = \pi_i^{1/2} P_{ij} 
\pi_j^{-1/2}$ is symmetric, as trivially follows from the detailed-balance 
condition. The spectral theorem then assures that $A$ 
has only real eigenvalues and that its eigenvectors form an orthonormal basis. 
The transition matrix $P$ has the same eigenvalues as $A$, and closely related 
eigenvectors:
\begin{equation}
 \sum_{j \in \Omega} \underbrace{\pi_i^{1/2} P_{ij} \pi_j^{-1/2} }_{A_{ij}} 
x_j= \lambda x_i\quad \
 \Leftrightarrow \quad \ 
 \sum_{j \in \Omega} P_{ij} \underbrace{ \glc \pi_j^{-1/2} x_j 
\grc}_{\xtilde_j} = \lambda \underbrace{ \glc \pi_i 
^{-1/2}  x_i \grc}_{\xtilde_i}.
\label{equ:AssociatedSymmetric}
\end{equation}
The eigenvectors $\xtilde$ of $P$ must be multiplied with 
$\sqrt{\pi}$ to be mutually orthogonal. They provide a basis 
on which any initial probability distribution 
$\pit{0}$ can be expanded. An irreducible and aperiodic 
transition matrix (reversible or not) has one eigenvalue $\lambda_1 = 1$, and 
all others satisfy $|\lambda_k| < 1 \ \forall k \ne 1$.

A non-reversible transition matrix may belong to a mix of three different 
classes, and this variety greatly complicates their mathematical analysis. $P$  
may have only real eigenvalues and real-valued eigenvectors, but without there 
being an associated symmetric matrix $A$, as in \eq{equ:AssociatedSymmetric} 
(see~\cite[Sect. 2.3]{Weber2017} for an example). A non-reversible transition 
matrix may also have only real eigenvalues but with geometrical 
multiplicities that not all agree with the algebraic multiplicities. 
Such a matrix is not diagonalizable. Finally, $P$ may have real and then pairs 
of complex eigenvalues~\cite{Weber2017}. This generic transition matrix can be 
analyzed in terms of its eigenvalue spectrum, and expanded in terms of a basis 
of eigenvectors~\cite{SakaiEigenvaluePRE2016,GwaSpohnPRL1992,Dhar1987}. As 
non-reversible transition matrices may well have only real eigenvalues, the 
size 
of their imaginary parts does not by itself indicate the degree of 
non-reversibility~\cite{Nielsen2015}. Reversibilizations of non-reversible 
transition matrices have been much studied~\cite{Fill1991}, but they modify the 
considered transition matrix. \\

\subsubsection{Total variation distance, mixing times}
\label{subsec:TVD}

In real-world applications, irreducibility and aperiodicity of a Markov chain 
can usually be 
established beyond doubt. The stationary distribution $\pi$ of a transition 
matrix constructed to satisfy a global-balance condition is also known 
explicitly. The 
time scale for the exponential approach of $\pit{t}$ to 
$\pi$---that always exists---is much more difficult to establish.
In principle, the difference between the two distributions is
quantified through the total variation distance:
\begin{equation}
    \tvd{\pit{t} - \pi} =  \max_{A \subset \Omega} | \pit{t}(A) - \pi(A) | = 
    \half \sum_{i \in \Omega} | \pit{t}_i - \pi_i|.
\label{equ_TotalVariation}
\end{equation}
The distribution $\pit{t}$ depends on the initial distribution, but for any 
choice of $\pit{0}$, for an irreducible and aperiodic transition matrix, the 
total variation distance  is smaller than an exponential bound $C \alpha^t$ 
with 
$\alpha \in (0,1)$ (see the convergence theorem for Markov chains~\cite[Theorem 
4.9]{Levin2008}). At the mixing time, the distance from the most unfavorable 
initial configuration,
\begin{equation}
    d(t) = \max_{\pit{0}} \tvd{\pit{t}(\pit{0}) - \pi}, 
\label{equ:TVD14}    
\end{equation}
drops below a certain value $\epsilon$
\begin{equation}
 \tmix(\epsilon) = \min\SET{t: d(t)\le \epsilon}. 
\end{equation}
Usually, $\epsilon = \tfrac 14$ is taken, with $\tmix = 
\tmix(1/4)$~\cite{Levin2008}. Here, the value of $\epsilon$ is arbitrary, but 
smaller than $1/2$ in order for convergence in a small part $A$  of $\Omega$ 
(without exploration of $\Omega \setminus A$) not to be counted as a success 
(an example is given in \subsect{subsec:VShape}). Once such a value smaller 
than $\tfrac 12$ is reached, exponential convergence in $t / \tmix $ 
sets in~\cite[eq. (4.36)]{Levin2008}. The mixing time is thus needed to 
obtain a first sample of the stationary distribution from a most unfavorable 
initial configuration. It does not require the existence of a spectrum of the 
transition matrix $P$, and it can be much larger than the correlation time 
$\tcorr$ given by the absolute inverse spectral gap of $P$, if it exists (the 
time it takes to decorrelate samples in the stationary regime). Related 
definitions of the mixing time take into account such thermalized initial 
configurations~\cite{Lovasz1998}. \\

\subsubsection{Diameter bounds, bottleneck, conductance}
\label{subsec:conductance}

An elementary lower bound for the mixing time on a graph $G = (\Omega, E)$ 
(with 
the elements of $\Omega$ as vertices, and the non-zero elements of the 
transition matrix as edges) is given by the graph diameter $L_G$, 
that is, the minimal number of steps it takes to connect any two vertices $i, j 
\in \Omega$. The mixing time, for any $\epsilon < 1/2$, trivially satisfies
\begin{equation}
 \tmix   \ge \frac {L_G}{2}.
\end{equation}
For the Metropolis or heatbath single-spin flip dynamics in the Ising model 
with $N$ spins (in any dimension $d$), mixing times throughout the 
high-temperature phase are logarithmically close to the diameter bound with the 
graph diameter $L_G=N$, the maximum number of spins that differ between $i$ and 
$j$~\cite{Martinelli1999}.

Mixing and correlation times of a MCMC algorithm can become very large if there 
is a bottleneck in the sample space $\Omega$ from which it is difficult 
to excape. Two remarkable insights are that in MCMC there is but a single such
bottleneck (rather than a sequence of them), and that the mixing time 
is bracketed (rather than merely bounded from below) by functions of the 
conductance~\cite{Chen1999}. Also  called \quot{bottleneck 
ratio}~\cite{Levin2008}, the conductance is defined as the flow across the 
bottleneck:
\begin{equation}
    \Phi\equiv
    \min_{S \subset\Omega, \pi_S \le \half}
    \frac{\FCAL_{S\to \Sbar}}{\pi_{S} }
   =
    \min_{S\subset\Omega, \pi_S \le \half }
    \frac{\sum_{i\in S, j \in \Sbar }\pi_{i} P_{ij} }{\pi_S }, 
\label{equ:DefConductance}
\end{equation}
where $\Sbar = \Omega \setminus S$.
Although it can usually not be computed in real-world applications, the 
conductance is of importance because the liftings which are at the heart of 
ECMC leave it unchanged.

For reversible Markov chains, the correlation time is bounded by the 
conductance as~\cite{Chen1999}: 
\begin{equation}
 \frac{1}{\Phi} \leq \tcorr \leq \frac{8}{\Phi^2}.
\label{equ:CorrelationConductance}
\end{equation}
The lower bound follows from the fact that to cross from $S$ into $\Sbar$, the 
Markov chain must pass through the boundary of $S$, but this  
cannot happen faster than through direct sampling within $S$, that is, 
with probability $\pi_i/\pi_S$ for $i$
on the boundary of $S$. The upper bound was proven in~\cite[Lemma 
3.3]{Sinclair1989}. For arbitrary MCMC, one has the relation
\begin{equation}
 \frac{1}{4\Phi} \leq \ACAL \leq \frac{20}{\Phi^2}, 
\label{equ:SetTimeConductance} 
\end{equation}
where $\ACAL$ is the \quot{set time}, the maximum over all sets $S$ of the 
expected time to hit a set $S$ from a configuration sampled from 
$\pi$, multiplied with the stationary distribution $\pi(S)$. In addition, the 
mixing time, defined in~\cite{Chen1999} with the help of a stopping rule, 
satisfies:
\begin{equation}
\frac{\const}{\Phi} < \tmix < \frac{\const'}{\Phi^2} \loga{\frac{1}{ \pi_0}}
\label{equ:MixingConductance}
\end{equation}
where the constants are different for reversible and for non-reversible Markov 
chains. 

The above inequalities strictly apply only to finite Markov chains, where the 
smallest weight $\pi_0$ is well-defined. A continuous system may have to be 
discretized in order to allow a discussion in its terms. Also, in the 
application to ECMC, which is event-driven, mixing and correlation times may 
not reflect computational effort, which roughly corresponds to the number of 
events, rather than of time steps. Nevertheless, the conductance yields the 
considerable diffusive-to-ballistic speedup that may be reached by 
non-reversible liftings~\cite{Chen1999} if the collapsed Markov chain is itself 
close to the $\sim 1 / \Phi^2$ upper bound of 
\eqtwo{equ:CorrelationConductance}{equ:MixingConductance}. \\

\subsection{Factorization}
\label{subsec:Factorization}

In generic MCMC, each proposed move is accepted or rejected based on the change 
in total potential that it entails. For hard spheres, a  move in the Metropolis 
algorithm~\cite[Chap. 2]{SMAC} can also be pictured as being accepted \quot{by 
consensus} (no pair of spheres presenting an overlap) and rejected \quot{by 
veto} (at least one overlap between two spheres). The \quot{potential 
landscape} 
and the \quot{consensus--veto} interpretations are here equivalent as the pair 
potential of two overlapping hard spheres is infinite, and therefore also the 
total potential.

The factorized Metropolis filter~\cite{Michel2014JCP,Faulkner2018} 
generalizes the  \quot{consensus--veto} picture to an arbitrary system
whose stationary distribution breaks up into a set $\MCAL$ 
of factors $M = \FACTOR{I_M}{T_M}$. Here, $I_M$ is an index set and 
$T_M$ a 
type (or label), such as \quot{Coulomb}, \quot{Lennard-Jones}, \quot{Harmonic} 
(see~\cite[Sect. 2A]{Faulkner2018}). The total potential $U$ of a 
configuration $c$ then writes as the sum over factor potentials 
$U_M$ that only depend on the factor configurations $c_M$:
\begin{equation}
 U(\underbrace{\SET{\rvec_1 \TO \rvec_N}}_{c}) = \sum_{M \in \MCAL}
U_M (\underbrace{\SET{\rvec_i: i \in I_M}}_{c_M}),
\label{equ:PotentialFactorized}
\end{equation}
and the stationary distribution appears as a product over exponentials of 
factor potentials:
\begin{equation}
\pi(\Conf) =\expc{-\beta U(\Conf) } = 
\prod_{M \in \MCAL} \pi_M(\Conf_M) =
\prod_{M \in \MCAL}\expc{-\beta U_{M}(\Conf_M)}.
\label{equ:BoltzmannFactorized}
\end{equation}
Energy-based MCMC considers the left-hand side of \eq{equ:PotentialFactorized} 
and molecular dynamics its derivatives (the forces on particles). All 
factorizations $\MCAL$ are then equivalent. 
In contrast, 
ECMC concentrates on the right-hand side of \eq{equ:PotentialFactorized}, and
different factorizations now produce distinct algorithms. \\

\subsubsection{Factorized Metropolis algorithm, continuous-time limit}
\label{subsec:FactorizedMetropolis}

The Metropolis filter~\cite{Metropolis1953} is a standard algorithm for  
accepting a proposed move from configuration $\Conf$ to configuration 
$\NewConf$:
\begin{equation}
    \pMet(\Conf \to \NewConf) = \min\glc 1, \frac{\pi_{c'} }{ \pi_c } \grc
    =  \min\glc 1, \expb{-\beta \Delta U } \grc
                                 = \min\glc 1, \prod_{M \in \MCAL} \expb{-\beta
\Delta U_{M}}
    \grc ,
\label{equ:MetropolisProduct}
\end{equation}
where $\Delta U_{M} = U_{M}(\NewConf_M) - U_{M}(\Conf_M)$. The factorized 
Metropolis filter~\cite{Michel2014JCP} plays a crucial role in ECMC. It inverts 
the order of product and minimization, and it factorizes as its name 
indicates: 
\begin{equation}
    \pFact(\Conf \to \NewConf) = \prod_{M \in \MCAL} \min \glc 1,
\expb{-\beta \Delta
    U_{M}} \grc.
\label{equ:MetropolisFactorized}
\end{equation}
The factorized Metropolis filter 
satisfies detailed balance. This is  because, for a single factor, 
$\pFact$  reduces to $\pMet$ (which obeys detailed balance~\cite{SMAC}) and 
because the Boltzmann weight of \eq{equ:BoltzmannFactorized} and the factorized 
filter of \eq{equ:MetropolisFactorized} factorize along the same lines. The 
detailed-balance property is required for proving the correctness of ECMC 
(see~\cite[eq. (27)]{Faulkner2018}), although ECMC is never actually run in 
reversible mode.

The Metropolis filter $\pMet$ of \eq{equ:MetropolisProduct} is implemented by 
sampling a Boolean random variable:
\begin{equation}
    X^{\text{Met}}(\Conf \to \NewConf) =
\begin{cases}
\text{\quot{True}} \quad &\text{if}\ \ranb{0,1} <
\pMet (\Conf \to \NewConf)
\\
\text{\quot{False}} \quad &\text{else} , \\
\end{cases}
\label{equ:MetropolisBoolean}
\end{equation}
where \quot{True} means that the move $\Conf \to \NewConf$ is accepted.
In contrast, the factorized Metropolis 
filter appears as a conjunction of Boolean random variables:
\begin{equation}
    X^{\text{Fact}}(\Conf \to \NewConf)  = \bigwedge_{M \in \MCAL}
X_{M}(c_M \to c'_{M}).
\label{equ:MetropolisFactorizedBoolean}
\end{equation}
The left-hand side of \eq{equ:MetropolisFactorizedBoolean} is \quot{True} if 
and only if all the independently sampled Booleans $X_{M}$ on the right-hand 
side are \quot{True}, each with probability $\min \glc 1, 
\expb{-\beta \Delta U_{M}} \grc$.
The above conjunction thus 
generalizes the consensus--veto picture from hard spheres to general 
interactions, where it is inequivalent to the energy-based filters. 

In the continuous-time limit, for continuously varying potentials, 
the acceptance probability of a factor $M$ becomes:
\begin{equation}
\min \glc 1, \expb{-\beta \Delta U_M} \grc =
\expb{-\beta \maxZeroa{\Delta U_M}}
\xrightarrow{\Delta U_M \to \diff U_M}
1 - \beta \maxZeroa{\diff U_M},
\label{equ:AcceptanceFactors}
\end{equation}
where
\begin{equation}
\maxZeroa{x} =  \max(0, x) \quad \forall x \in \RR
\label{equ:UnitRamp}
\end{equation}
is the unit ramp function.  The acceptance probability of the 
factorized Metropolis filter then becomes
\begin{equation}
    \pFact(\Conf \to \NewConf) = 1 - \beta \sum_{M \in \MCAL}
\maxZeroc{\diff
    U_M (\Conf_M \to \NewConf_M) },  
\label{equ:MetropolisFactorizedInfinitesimal}
\end{equation}
and the total rejection probability for the move turns into a sum over factors:
\begin{equation}
1 - \pFact (\Conf \to \NewConf)  = \beta\sum_{M \in \MCAL} \maxZeroc{\diff
    U_M (\Conf_M \to \NewConf_M)}.
\label{equ:MetropolisRejectionInfinitesimal}
\end{equation}
In ECMC, the infinitesimal limit is used to break possible ties, so that a veto 
can always be attributed to a unique factor $M$ and then transformed into a 
lifting. The MCMC trajectory in between vetoes appears deterministic, and the 
entire trajectory 
as piecewise deterministic~\cite{Bierkens2017,BierkensPDMC2017}
(see also~\cite{Michel2016PHD}). In the event-driven formulation 
of ECMC, rather than to check consensus for each move $c \to c'$, factors 
sample candidate events (candidate vetoes), looking ahead on the
deterministic trajectory. The earliest candidate event is 
then realized as an event~\cite{Peters_2012}. \\

\subsubsection{Stochastic potential switching}
\label{subsec:StochasticPotentialSwitching}
Whenever the factorized Metropolis filter signals a consensus, the system 
appears as non-interacting, while a veto makes it appear hard-sphere like. This 
impression is confirmed by a mapping of the factorized Metropolis filter onto a 
hamiltonian that stochastically switches factor potentials~\cite{Mak2005}, in 
our case between zero and infinity.

In stochastic potential switching, a factor potential  $U_M$ 
(see~\cite[Sect. V]{Mak2005}) for a move $c \to c'$ is 
switched to $\Utilde_M$ with probability
\begin{equation}
  S_M(c^\dagger) = \expd{ \beta \glc U_M(c^\dagger) - \Utilde_M(c^\dagger) -   
\Delta U_M^*(c, 
c')\grc},
\label{equ:SwitchingSM}  
\end{equation}
with  $c^\dagger = c$. The switching affects both configurations, but is 
done with 
probability $S_M(c)$~\cite{Mak2005}. The constant $\Delta 
U_M^*$ is chosen so that 
$S_M < 1$ for both $c$ and $c'$, for example as $\Delta U_M^*= 
\max\glc U_M(c), U_M(c') \grc + \epsilon$ with $\epsilon \gtrsim 0$. 
If the potential is not switched, the move $c \to c'$ is subject to a 
pseudo-potential
\begin{equation}
  \Ubar_M(c^\dagger) = U_M(c^\dagger) - \beta^{-1} \log [1-S_M(c^\dagger)]
\label{equ:SwitchingUBar}  
\end{equation}
for both configurations $c^\dagger \in \SET{c, c'}$. 
ECMC considers \quot{zero-switching} towards 
$\Utilde(c) = 
\Utilde(c') =0$. If $U_M(c') < U_M(c)$, the zero-switching probability is $ 
\sim 1 - \beta \epsilon \to 1$ for $\epsilon \to 0$.

If $U_M(c') > U_M(c)$, the zero-switching probability 
is smaller than $1$. If the potential is not switched so that the
pseudo-potential $\Ubar$ is used at $c$ and $c'$, 
then 
$\Ubar_M(c)$ remains finite while $\Ubar_M(c') \sim - \logb{\beta\epsilon}/ 
\beta \to \infty$ for $\epsilon \to 0$. 
In that limit, the pseudo-potential barrier $\Ubar(c') - \Ubar(c) $ 
diverges towards the hard-sphere limit.
For $U_M(c') > U_M(c)$ with ($\glc U_M(c') - U_M(c) \grc \to \diff 
U_M > 0$, the zero-switching probability approaches 
$\sim 1 - \beta \diff U_M$ and, together with the case $U_M(c') < U_M(c)$, 
this yields
\begin{equation}
 S_M(c) \to 1 - \beta \glc \diff U_M(c \to c')\grc^+, 
\end{equation}
with the unit ramp function of \eq{equ:UnitRamp}.  In the infinitesimal limit
$U_M(c') - U_M(c) \to \diff U_M \ \forall M \in \MCAL$, 
at most one factor fails to zero-switch. The factorized 
Metropolis filter of \eq{equ:MetropolisFactorizedInfinitesimal} follows 
naturally. \\

\subsubsection{Potential invariants, factor fields}
\label{subsec:FactorFields}

Invariants in physical systems originate in conservation laws and topological 
or 
geometrical characteristics, among others. Potentials $V$ that are invariant 
under ECMC time evolution play a special role if they can be expressed as a sum 
over factor potentials $V_M$. The total system potential  then writes as 
$\Utilde = U + V$ with constant $V = \sum_{M \in \MCAL}  V_M$, which results in
\begin{equation}
\Utilde = \sum_{M \in \MCAL} \Utilde_{M} = \sum_{M \in \MCAL} \glb 
U_M + V_M \grb.
\end{equation}
Although $V$ is constant, the factor terms $V_M$ can vary between 
configurations. In energy-based MCMC and in molecular dynamics, such constant 
terms play no 
role. \subsect{subsec:FactorFieldsOneD} reviews linear factor invariants   
that drastically reduce mixing times and dynamical critical exponents
in one-dimensional particle systems. \\

\subsection{Lifting}
\label{subsec:Lifting}

Generic MCMC  proposes moves that at each time are randomly sampled from a 
time-independent set. Under certain conditions, the same moves can be proposed 
in a less-random order without changing the stationary distribution. 
Lifting allows one to formulate the resulting 
algorithm as a random process without memory, that is, a Markov chain. 
\secttwo{sec:SingleParticles}{sec:OneDimensionalN} will later review examples 
of 
MCMC, including ECMC, where lifting substantially reduces mixing and 
autocorrelation times and their scaling with system size.

Mathematically, a lifting of a Markov chain (with its opposite being a 
\quot{collapsing}) consists in 
a mapping $f$ from  a \quot{lifted} sample space $\Omegahat$ to 
the \quot{collapsed} one, 
$\Omega$, such that each configuration $v\in \Omega$ splits
(\quot{lifts}) into copies $i \in f^{-1}(v)$.
There are two requirements ~\cite[Sect. 3]{Chen1999}. The total stationary 
probability of all lifted copies must 
equal the original stationary probability:
\begin{equation}
 \pi_v = \pihat\glc f^{-1}(v) \grc = \sum_{i \in f^{-1}(v)} \pihat_i. 
\label{equ:LiftedStationaryWeights}
\end{equation}
Also, the sum of the flows between the lifted copies of two original 
configurations $u$ and $v$ must equal the flow between the collapsed 
configurations:
\begin{equation}
\underbrace{\pi_v P_{vu}}_{\text{collapsed flow}} = \sum_{i \in f^{-1}(v), j 
\in f^{-1}(u)} 
\overbrace{\pihat_i \Phat_{ij}}^{\text{lifted flow}}.
\label{equ:LiftedTransitionMatrix}
\end{equation}
ECMC usually considers a special lifted sample space $\Omegahat = \Omega 
\times \LCAL$, with $\LCAL$ a set of lifting variables $\sigma$. The liftings 
$i$ of a collapsed configuration $v$ then write as $i = (v,\sigma)$, so that 
each collapsed configuration has the same number of copies. In addition, the 
lifting variables usually are without influence on the stationary distribution 
so that: 
\begin{equation}
     \frac{\pihat(u,\sigma)}{ \pi(u)} = 
\frac{\pihat(v,\sigma)}{\pi(v)}\quad \forall\ u, v \in \Omega; \forall\ \sigma 
\in \LCAL.
\end{equation}

A lifted Markov chain  cannot have a larger conductance than its collapsed 
counterpart, because the statistical weights and the flows in 
\eqtwo{equ:LiftedStationaryWeights}{equ:LiftedTransitionMatrix} remain 
unchanged, and therefore also the 
bottlenecks in \eq{equ:DefConductance}. Also, from 
\eq{equ:LiftedTransitionMatrix}, the reversibility of the lifted Markov chain 
implies reversibility of the collapsed chains. In this case, lifting can only 
lead to marginal speedup~\cite{Chen1999}.  Conversely, a non-reversible 
collapsed Markov chain corresponds to a non-reversible lifted Markov chain. 
Within the bounds of 
\eq{equ:SetTimeConductance} and the corresponding expression for 
mixing times, 
the lifted Markov chain may be closer to the $\sim 1/\Phi$ lower 
\quot{ballistic} mixing-time limit, for the collapsed Markov chain at the $\sim 
1/\Phi^2$ upper \quot{diffusive} limit.  \\

\subsubsection{Sequential MCMC, particle lifting}
\label{subsec:SequentialMC}

The sequential version of Metropolis MCMC for $N$-particle systems consists in 
performing single-particle moves with a consecutive rather than a 
random order of particle indices. 
It was proposed in 1953, in the founding publication of MCMC, which states
\quot{\dots \emph{we move each of the particles 
in succession} \dots}~\cite[p.22]{Metropolis1953JCP}.
Sequential MCMC is a non-reversible \quot{particle} 
lifting of the Metropolis algorithm or any other reversible Markov chain 
with single-particle moves, with a lifted sample space $\Omegahat = \Omega 
\times \SET{1 \TO N}$, where $\Omega = \SET{ \xvec = (x_1 \TO x_N)}$ is the 
collapsed sample space of particle positions.
With the lifted configurations 
$(\xvec, k)$, with $k$ the active particle,  and with stationary distribution 
$\pihat(\xvec, k) = \pi(\xvec)/N $,
the lifted transition matrix is
\begin{equation}
\Phat_{(\xvec,k),(\xvec',l)} = N P_{\xvec, \xvec'} 
\delta(x_1, x'_1) \TO \delta(x_{k-1},x'_{k-1})
\delta(x_{k+1}, x'_{k+1}) \TO \delta(x_N, x'_N) \delta_{k+1, l}
\label{equ:LiftingSequential}
\end{equation}
where the periodicity in the indices is understood. The factor 
$N$ on the right-hand side of \eq{equ:LiftingSequential} compensates for the 
\emph{a priori} probability of choosing the active particle $k$, which 
is $\tfrac 1N$ in the collapsed MCMC and $1$ in the lifted MCMC. Sequential 
MCMC satisfies the global-balance condition if the collapsed Markov chain is 
reversible (see \subsect{subsec:NonRevSeqForward} for an incorrect sequential 
version of a non-reversible MCMC). \\

In particle systems with central potentials, sequential MCMC is found to be 
slightly  more efficient than its collapsed reversible 
counterpart~\cite{OKeeffe2009,KapferKrauth2017,Lei2018_OneD}. Likewise, in the 
Ising model and related systems, the analogous spin sweeps (updating
spin $i+1$ after spin $i$, \emph{etc}.) again appear 
marginally faster than the random sampling of spin indices~\cite{Berg2004book}. 
Sequential MCMC is but a slightly changed variant of a reversible Markov chain, 
yet its particle lifting, the deliberative choice of the active particle, is 
one of the key features of ECMC. \\

\subsubsection{Displacement Lifting}
\label{subsec:DisplacementLifting}

In the local MCMC algorithms which are the collapsed counterparts of ECMC, the 
proposed moves can (roughly) be characterized by the independent sampling of a 
particle (or spin) $i \in \SET{1 \TO N}$ and of a displacement $\deltavec$ from 
a displacement set $\DCAL$. The \quot{displacement lifting} refers to a 
less-random sampling of the elements of $\DCAL$ in the lifted transition 
matrix. For a single particle, on a one-dimensional path graph $P_n$ 
displacement lifting, with $\DCAL = \SET{-1,+1}$ corresponding to  positive and 
to negative hops so that $\Omegahat = \Omega \times \DCAL$), is reviewed in 
\subsect{subsec:Diaconis} (see~\cite{Diaconis2000}). The lifting scheme, the 
choice of the lifted transition matrix, strives to preserve $\deltavec$ from 
one move to the next as much as is compatible with global balance and 
aperiodicity. Displacement lifting applies in more than one dimensions, for 
example for systems of dipole particles~\cite{Qin2020fastsequential}, whose 
dynamics is profoundly modified with respect to that of its collapsed 
counterpart. The displacement lifting in ECMC repeatedly uses the same 
infinitesimal displacement. \\

\subsubsection{Event-chain Monte Carlo}
\label{sec:ECMC}

ECMC~\cite{Bernard2009,Michel2014JCP}, in its use of factorized potentials and 
the factorized Metropolis algorithm, realizes the extreme limit of particle 
lifting and of displacement lifting. If compatible with global 
balance 
and irreducibility, it reutilises its moves, that is, keeps the active 
particles and their displacements from one time step to the next. ECMC moves 
are thus persistent, and their randomness is strongly reduced, rendering its 
trajectories piecewise deterministic (compare with 
\subsect{subsec:FactorizedMetropolis}). The persistence of moves requires the 
global-balance condition to be carefully checked at every infinitesimal time 
step.

In ECMC, factorization and the consensus principle of the factorized Metropolis 
algorithm ensure that a single factor triggers the veto that terminates the 
effectively non-interacting trajectory (as discussed in
\subsect{subsec:StochasticPotentialSwitching}). This factor is sole responsible 
for the ensuing lifting. The triggering factor has at least one active particle 
in its in-state. The out-state can be chosen to have exactly the same number of 
active elements. In translationally invariant particle systems (and in 
similarly invariant spin systems), the lifting does not require a change of 
displacement vector $\deltavec$, and concerns only the particle sector of the 
lifting set $\LCAL$. For a two-particle factor, the lifting scheme is 
deterministic, and the active and the passive particle interchange their roles. 
For a three-particle factor, the active particle may have to be sampled from a 
unique probability distribution, so that the lifting scheme is stochastic but 
uniquely determined~\cite{Harland2017}. For larger factors, the next active 
particle may be sampled following many different stochastic lifting schemes 
that can have inequivalent dynamics~\cite{Harland2017} (see also~\cite[Sect. 
II.B]{Faulkner2018}). \\

\subsection{Thinning}
\label{subsec:Thinning}

Thinning~\cite{LewisShedler1979,Devroye1986} denotes a strategy where an event 
in a complicated time-dependent random process (here the decision to veto a 
move) is first provisionally established on the basis of an approximate 
\quot{fatter} but simpler process, before being thinned (confirmed or 
invalidated) in a second step. Closely related to the rejection method in 
sampling~\cite{SMAC}, thinning doubly enables all but the simplest ECMC 
applications. First, it is used when the factor event rates $\beta 
\maxZeroc{\diff U_M}$ of \eq{equ:MetropolisRejectionInfinitesimal} cannot be 
integrated easily along active-particle trajectories, while 
appropriate bounding potentials allow integration. Second, thinning underlies 
the cell-veto 
algorithm~\cite{KapferKrauth2016}, which ascertains consensus and identifies 
vetoes among $\sim N$ factors (the number of particles) with an effort that 
remains independent of $N$. The random process based on the untrimmed bounding 
potential is time-independent and pre-computed, and it can be sampled very 
efficiently.

Thinning thus uses approximate bounding potentials for the bulk of the 
decision-making in ECMC while, in the end, allowing the sampling to remain 
approximation-free. This contrasts with molecular dynamics, whose numerical 
imperfections---possible violations of energy, momentum, and 
phase-space conservation under time-discretization---cannot be fully 
repaired (see~\cite{Zhong1988}). ECMC also eliminates the rounding errors for 
the total potential, that cannot be avoided in molecular dynamics for
long-range Coulomb interactions. \\

\subsubsection{Thinning and bounding potentials}
\label{subsec:Bounding}

Thinning is unnecessary for hard spheres and for potentials with simple 
analytic expressions that can be integrated analytically. In other cases, it 
may be useful to compare a factor event rate to a (piecewise) upper bound, as 
$\maxZeroc{\beta \diff U_M} \le q_M^{\max}$. In the case where, for simplicity, 
the active particle $k$ moves along the $x$ direction, the bounding factor 
event rate allows one to write
\begin{equation}
\maxZeroc{\beta \diff U_M} = \beta \frac{\partial U_M^+}{\partial x_k} \diff 
x_k = \underbrace{q_M^{\max} \diff x}_{\text{constant-rate}} \underbrace{\frac{ 
\partial U_M^+ / \partial x_k}{q_M^{\max}}}_{\text{confirmation}}.
\end{equation}
The factor $M$ triggers the veto with the factor event rate on the left-hand 
side of this equation, which varies with the configuration. On the right-hand 
side, this is expressed as a product of independent probabilities corresponding 
to a conjunction of random variables, of which the first corresponds to a 
constant-rate Poisson process and the second to a simple rejection rate. The 
second factor must be checked only  when the first one is \quot{True}. Thinning 
thus greatly simplifies ECMC when the mere evaluation of $U_M$ and its 
derivatives is time-consuming. The JeLLyFysh application of 
\subsect{subsec:Jellyfysh} implements thinning for potentials in a variety of 
settings (see~\cite{Hoellmer2020}). \\

\subsubsection{Thinning and the cell-veto algorithm, Walker's algorithm}
\label{subsec:Walker}

The cell-veto algorithm~\cite{KapferKrauth2016} (see 
also~\cite{BouchardCote2018}) permanently tiles  two or three-dimensional 
sample space (with periodic boundary conditions) into cells with precomputed 
cell bounds for the factor event rates of  any long-ranged factor that touches 
a pair of these cells. In this way, for example, a long-range Coulomb event 
rate  for a pair of distant atoms (or molecules) can be bounded by the cell 
event rate of the pair of cells containing these atoms. For an active atom in a 
given cell, the total event rate due to other far-away atoms can also be 
bounded 
by the sum over all the cell event rates. Identifying the vetoing atom 
(factor) then requires to provisionally sample a cell among all those 
contributing to the total cell event rate. This problem can be solved in 
\bigObs{1} using Walker's 
algorithm~\cite{Walker1977AnEfficientMethod,Devroye1986}.
After identification of the 
cell, the 
confirmation step involves at most a single factor. The cell-veto algorithm is 
heavily used the JeLLyFysh application (see \subsect{subsec:Jellyfysh}). In 
essence, the cell-veto algorithm thus establishes consensus and identifies a 
veto among $\sim N$ factors in \bigObs{1} in a way that installs neither 
cutoffs 
nor discretizations.  \\

\section{Single particles on a path graph}
\label{sec:SingleParticles}  

The present section reviews liftings for one-particle MCMC on a path graph 
$P_n= (\Omega_n, E_n)$, where the sample space is $\Omega_n = \SET{1 \TO n}$, 
and where the set of edges $E_n= \SET{(1,2) \TO (n-1,n)}$ indicates the 
non-zero 
elements of the transition matrix. The graph $P_n$ thus forms a one-dimensional 
$n$-site lattice without periodic boundary conditions. The stationary 
distribution is $\SET{\pi_1 \TO \pi_n}$. 
Two virtual vertices $0$ and $n+1$ and virtual 
edges $(0,1)$ and $(n,n+1)$,
with $\pi_0 = \pi_{n+1} = 0$,
avoid the separate discussion of the boundaries. The 
Metropolis algorithm on $P_n$ (with the virtual additions, but an unchanged 
sample space) proposes a move from vertex $i \in \Omega_n$ to $j = i \pm 1$ 
with probability $\tfrac 12$. This move is accepted with the Metropolis 
probability $\min(1, \pi_j/\pi_i)$. If rejected, the particle remains at $i$. 
The equilibrium flows:
\begin{equation}
  \begin{tikzcd}[column sep = huge]
\framebox{i-1} \arrow[yshift = 0.5 ex]{r}{\half \min(\pi_{i-1},\pi_i)}& 
 \arrow[loop above]{}{\underbrace{\pi_i - \tfrac12 \min(\pi_{i-1},\pi_i) - 
\tfrac12 \min(\pi_{i},\pi_{i+1})}} \framebox{i}
\arrow[yshift = 0.5 ex]{r}{\half \min(\pi_i,\pi_{i+1})}  
\arrow[yshift = - 0.5 ex]{l}{\half \min(\pi_i,\pi_{i+1})}  
& \framebox{i+1}  \arrow[yshift = -0.5 ex]{l}
{ \half \min(\pi_i,\pi_{i+1}) }
\end{tikzcd} 
\label{equ:LiftedTwoParticlesDetailed}
\end{equation}
satisfy detailed balance, and the total flow into each configuration $i$ 
(including from $i$ itself)
satisfies the global-balance condition of \eq{equ:GlobalBalance}, as already 
follows from detailed balance.

The displacement-lifted Markov chains in this section, with $\Omegahat_n= 
\Omega_n \times \SET{-1,+1}$, split the Metropolis moves $i \to i \pm 1$ into 
two copies (the virtual vertices and edges are also split). The lifted MCMC 
algorithm
consists in the succession of a transport transition matrix 
$\Ptrans$ and a re-sampling 
transition matrix $\Pres$.
$\Ptrans$
proposes a move from 
vertex $(i, \sigma)$ to vertex $(i+ \sigma,\sigma)$ with 
probability $1$ and  accepts with the Metropolis filter $\min(1, 
\pi_{i+ \sigma}/\pi_{i})$ given that
$\pihat_{(k,\sigma)} = \half \pi_k \ \forall k$.
In the \quot{$+1$} copy, this 
corresponds to forward moves, and in the \quot{$-1$} copy to backward moves. 
All rejections are replaced by liftings: If the move is not accepted, a 
\quot{lifting} move $(i, \sigma) 
\to (i, -\sigma)$ takes place.
$\Ptrans$, which is not a Metropolis 
filter, satisfies the global-balance condition for each 
lifted configuration, as the 
flow into $(i, \sigma)$ equals $\half \pi_i = 
\pihat_{(i, \sigma)}$:
\begin{equation}
\text{transport flows:}\quad
  \begin{tikzcd}[column sep = huge]
\framebox{(i-1,+1)} \arrow{r}{\half  \min(\pi_{i-1},\pi_i)}& \framebox{i,+1}
\arrow{r}{\half \min(\pi_i,\pi_{i+1})} \arrow[xshift=-0.5ex,swap]{d}{\half 
[\pi_i - \min(\pi_i,\pi_{i+1})]} & \framebox{i+1,+1} \\
   \framebox{i-1,-1} &\framebox{i,-1} \arrow{l}{\half \min(\pi_{i-1},\pi_i)} 
\arrow[xshift=.5ex,swap]{u}{\half [\pi_i -\min(\pi_{i-1},\pi_i)]}& 
\framebox{i+1,-1} \arrow{l}{\half \min(\pi_i,\pi_{i+1})} \end{tikzcd} ,
\label{equ:LiftedTwoParticlesTransport}
\end{equation}
In the re-sampling stage, the configuration $(i, \sigma)$ is simply switched to 
$(i, -\sigma)$ with a small probability $\epsilon = 1/n$. $\Pres$ satisfies 
detailed balance, and the re-sampling flows are: 
\begin{equation}
\text{re-sampling flows:}\quad
  \begin{tikzcd}[column sep = huge]
\framebox{(i,+1)}\arrow[xshift=-0.5ex,swap]{d}{\half 
\pi_i \epsilon} \\
  \framebox{(i,-1)} \arrow[xshift=.5ex,swap]{u}{\half \pi_i 
\epsilon}
 \end{tikzcd} .
\label{equ:LiftedTwoParticlesResampling}
\end{equation}
The lifted transition matrix $\Phat = \Ptrans \Pres$
satisfies global balance. In ECMC applications, the re-sampling 
stage after each transport stage is abandoned in favor of a carefully designed 
re-sampling after a large number of transport steps.

The speedups that can be achieved by lifting on the path graph $P_n$ depend on 
the choice of $\pi$. Bounded stationary distributions are reviewed in 
\subsect{subsec:BoundedOneDSingle}) and unbounded ones in 
\subsect{subsec:UnboundedOneDSingle}. All Markov chains on $P_n$ naturally 
share a diameter bound $n$ which is independent of $\pi$ and that (up to a 
constant) is the same for the lifted chains.  \\

\subsection{Bounded one-dimensional distributions}
\label{subsec:BoundedOneDSingle}

In real-world applications of MCMC, many aspects of the stationary distribution 
$\pi$ remain 
unknown.  Even the estimation of the  minimum or the maximum of $\pi_i$ is often
a non-trivial computational problem treated by simulated annealing, a variant 
of 
MCMC~\cite{Kirkpatrick1983,SMAC}. Also, crucial characteristics as the 
conductance are usually unknown, and the mixing time can at best be inferred 
from the running simulation. Simplified models, as a particle on a path graph, 
offer more control. This subsection reports on results for stationary 
distributions for  which $(\max_{i \in \Omega_n} \pi_{i} ) / (\min_{j \in 
\Omega_n} 
\pi_j) $ remains finite for $\Npath \to \infty$. The conductance bound then 
scales as $\Phi \sim n$ for $n \to \infty$, and reversible local MCMC mixes in 
\bigObs{n^2} steps. \\

\subsubsection{Constant stationary distribution, transport-limited MCMC}
\label{subsec:Diaconis}

The constant stationary distribution $\pi = \SET{ \tfrac 1n \TO \tfrac 1n }$ 
on $P_n$ (see~\cite{Diaconis2000}) illustrates the lifting concept. The 
collapsed Metropolis MCMC moves with probability $\half$ to the left and to the 
right. This Markov chain is aperiodic because of the rejections triggered by 
the aforementioned virtual vertices and edges (which are not counted in $P_n$). 
The collapsed Markov chain performs a diffusive random 
walk. Its \bigObs{n^2} mixing time is close to the upper mixing-time bound 
of \eq{equ:MixingConductance}. As $\tmix$  is on larger time scales than 
the lower conductance bound, the collapsed Markov chain appears \quot{transport 
limited}. Reorganizing its moves through lifting is thus promising.

In the lifted sample space $\Omegahat_n = \Omega_n \times \SET{+,-}$, the 
transition matrix $\Ptrans$ describes a deterministic rotation on the 
lifted graph which is equivalent to a ring with $2n$ vertices and periodic 
boundary conditions induced by the lifting moves  $(i=1, \sigma = -1) \to (1,  
1)$ and $(i=n, \sigma = +1) \to (n,  -1)$ at the boundaries (see 
\eq{equ:LiftedTwoParticlesTransport}). Randomness is introduced by the 
re-sampling with rate $\epsilon = \tfrac 1n$ with $\Pres$, which switches the 
copies. The mixing time of the combined transition matrix $\Ptrans \Pres$ is 
$\tmix = \bigObs{n}$ (see~\cite{Diaconis2000}).

The re-sampling of $\Pres$ is not the only strategy for preserving some 
randomness in the model. Restarts provide an alternative to small-probability 
re-sampling at each time step, where after a given number $\ell$ of moves with 
one type of lifting variables, the type is changed. Restarts can be formulated 
as Markov chains, with a lifting variable that performs a countdown. In ECMC, 
restarts are often required to ensure irreducibility. \\

\subsubsection{Square-wave stationary distribution, inefficient liftings}

On the path graph $P_n$ with the  square-wave stationary distribution 
$\pi_{2k-1} = \tfrac{2}{3n}$, $\pi_{2k} = \tfrac{4}{3n}\ \forall k \in \SET{1 
\TO n/2}$ (for even $n$), the conductance $\Phi = \tfrac{2}{3 n}$ for $n \to 
\infty$ is  of the same order of magnitude as for the flat distribution. Its 
bottleneck (the vertex where the conductance is realized), is again at $i= 
n/2$. The Metropolis algorithm proposes moves from vertex $i$ with 
probabilities $\tfrac12$  to $i \pm 1$ but rejects them with probability 
$\tfrac 12$ if $i$ is odd (see \eq{equ:MetropolisProduct}). Its mixing time is 
\bigObs{n^2}, on the same order as in
\subsect{subsec:Diaconis}. In the lifted sample space $\Omegahat = \Omega 
\times \SET{+,-}$, the transition matrix $\Ptrans$ generates lifting moves with 
probability $ \half$ on odd-numbered vertices, and the persistence of the 
lifting variable is lost. In consequence, the lifting $i \to (i, \pm 1)$ is 
inefficient for the square-wave stationary distribution, and the mixing time 
scale unchanged as $\tmix = \bigObs{n^2}$. An additional set of height-type 
lifting variables, that decompose $\pihat$ into a constant \quot{lane} and an 
another one that only lives on even sites, recovers \bigObs{n} 
mixing~\cite{HayesJanes2013}.  

Lifting on the path graph is thus more difficult to put into practice for the 
square-wave than for the homogeneous distribution, illustrating lifted 
mixing-time exponents are sensitive  to the details of $\pi$. ECMC in real-world 
applications is 
likewise simpler for homogeneous systems with a global translational invariance 
(or with a similar invariant for spin systems). Remarkably however, ECMC can 
apply the same rules to any continuous interparticle potential in homogeneous 
space. It achieves considerable speedup similar to what the lifting of 
\subsect{subsec:Diaconis} achieves for the flat stationary distribution on the 
path graph. \\
 
\subsection{Unbounded one-dimensional stationary distributions}
\label{subsec:UnboundedOneDSingle}

For unbounded ratios of the stationary distribution ($\max_{ij} (\pi_i / \pi_j) 
\to \infty$ for $n \to \infty$) the conductance limits as well as the benefits 
that can be reaped through lifting vary strongly. Exact results for a V-shaped 
distribution and for the Curie--Weiss model are reviewed.\\

\subsubsection{V-shaped stationary distribution, conductance-limited MCMC} 
\label{subsec:VShape}

The V-shaped stationary distribution on the path graph $P_n$ of even length $n$ 
is given by $\pi_i = \const | \tfrac{n+1}{2} -i | \ \forall i \in \SET{1 \TO 
n}$, where $\const = \tfrac{4}{\Npath^2}$ (see~\cite{Hildebrand2004}). The 
stationary distribution $\pi$ thus decreases linearly from $i=1$ to the 
bottleneck $i= \tfrac n2$, with $\pi(\tfrac n2) = \pi(\tfrac n2 + 1) = 
\tfrac{\const}{2} \sim n^{-2}$ and then increases linearly from $i = \tfrac n2 
+1$ to $i=n$. The Metropolis algorithm again proposes a move from vertex $i$ to 
$i\pm 1$ with probabilities $\tfrac 12$, and then accepts them with the filter 
of \eq{equ:MetropolisProduct} (with the virtual end vertices ensuring 
the correct treatment of boundary conditions). The conductance equals 
$\Phi= \tfrac{2}{n^2}$, for the minimal subset $S = \SET{ 1 \TO n/2}$ (see 
\eq{equ:DefConductance}). The Metropolis algorithm mixes in $S$ on an 
\bigObs{n^2} diffusive timescale, but requires \bigObs{n^2 \log n} time steps 
to mix in $\Omega$~\cite{Hildebrand2004}. However, even a direct sampling in 
$S$, that is, perfect equilibration, samples the boundary vertex $n/2$ only 
on a  $\pi_i/\pi_S \sim n^{-2}$ inverse time scale. For the V-shaped 
distribution, the benefit of lifting is thus at best marginal 
(from \bigObs{n^2 \log n} to \bigObs{n^2}), as the collapsed Markov chain is 
already conductance-limited, up to a logarithmic factor.

The optimal speedup for the V-shaped distribution is indeed realized with 
$\Omegahat = \Omega \times \SET{+,-}$, and the transition matrices $\Ptrans$ 
and $\Pres$. The lifted Markov chain reaches a total variation distance of 
$\tfrac 12$ and  mixes in $S$ on an \bigObs{n} timescale. A total variation 
distance of $\epsilon < \tfrac 1 2$, and mixing in $\Omegahat$ is reached in 
\bigObs{n^2 } steps only~\cite{Hildebrand2004}.  This illustrates that 
$\epsilon <  \frac 12$ is required in the definition of \eq{equ:TVD14}. \\

\subsubsection{Curie--Weiss model, mixing times vs correlation times}

The Curie--Weiss model (or mean-field Ising model~\cite{Baxter1982}) for $N$ 
spins $s_i = \pm 1$ has the particularity that its total potential
\begin{equation}
 U = - \frac{J}{2N}  \sum_{i,j} s_i s_j = - \frac{J}{2N} M^2
\end{equation}
only depends on the magnetization $\sum_i s_i = M$. The mapping $k = (M + N ) / 
2 + 1$, $n = N + 1$, places it on the path graph $P_n$. The heat-bath algorithm 
(where a single spin is updated) can be interpreted as the updating of the 
global magnetization state. The model was studied in a number of 
papers~\cite{Levin2010,Turitsyn2011,FernandesWeigelCPC2011,Bierkens2017}.

The conductance bound of \eq{equ:DefConductance} for the Curie--Weiss model 
derives from a bottleneck at $k = n/2$. In the paramagnetic phase, one 
has $\Phi  = \bigObs{n^{1/2}}$, and at the critical point $\beta J=1$, $\Phi = 
\bigObs{ n^{3/4}}$ (see~\cite{Levin2010}). These bounds are smaller than $n$ 
because the magnetization is strongly concentrated around $M=0$ (that is,  $k = 
n/2$), in the paramagnetic phase. In the ferromagnetic phase, $\beta J >1$, the 
magnetizations are concentrated around the positive and negative magnetizations 
($k\gtrsim 1$ and $k \lesssim n$) and in consequence $\Phi$ grows exponentially 
with $n$, and so do all mixing times. 

The reversible Metropolis algorithm mixes in $n \log n$ steps for $T > T_c$ 
(see~\cite{Levin2010}). This saturates the upper bound of 
\eq{equ:MixingConductance}. At the critical point, the mixing time is as 
$n^{3/2}$ at $T=T_c$. As is evident from the small conductance bound, the 
Metropolis algorithm is again limited by slow diffusive transport and not by a 
bottleneck in the potential landscape. 

The lifted algorithm improves mixing times in both cases to the optimal. The 
correlation time can be shorter than $ n$ because of the  strong concentration 
of the probability density on a narrow range of 
magnetizations~\cite{Turitsyn2011,FernandesWeigelCPC2011,Bierkens2017}, while 
mixing time on the path graph is always subject to the $\sim n$ diameter bound. 
By contrast, in the paramagnetic phase $J > 1$, the conductance bound is 
exponential in $n$, and all mixing times are limited by the central bottleneck 
and not by slow transport. \\

\section{N particles in one dimension}
\label{sec:OneDimensionalN}

The present section reviews MCMC for $N$ hard-sphere particles
on a one-dimensional graph with periodic boundary conditions 
(the path graph $P_n$ with an added edge $(n, 1)$), and on continuous intervals 
of length $L$ with and without periodic boundary conditions. In all cases, 
moves respect the fixed order of particles ($x_1 < x_2 \TO < x_{N-1} < x_N$, 
possibly with periodic boundary conditions in positions and particle indices). 
With periodic boundary conditions, uniform rotations of the configuration are 
ignored, as they mix very slowly~\cite{RandallWinklerCircle2005}). The 
hard-sphere MCMC dynamics is essentially independent of the density for the 
discrete cases (see~\cite[eq. (2.18]{Lacoin2016detailed}) as well as in the 
continuum (see~\cite{KapferKrauth2017} and~\cite[Fig. 1]{Lei2018_OneD}). Most 
of the hard-sphere results are numerically found to generalize to a class of 
continuous potentials (see~\cite[Supp. Item 5]{KapferKrauth2017}). 

\subsect{subsec:NPartOneDimReversible} reviews exact mixing and 
correlation-time results for reversible Markov chains and 
\subsect{subsec:NPartOneDimNonReversible} those for non-reversible ones, 
including the connection with the totally asymmetric simple exclusion model 
(TASEP). \subsect{subsec:NPartOneDimLifted} discusses particle-lifted Markov 
chains, like the lifted TASEP, the lifted-forward Metropolis algorithm as well 
as ECMC, for which rigorous mixing times were obtained. In many cases, 
non-reversible, and in particular non-reversible lifted Markov chains and ECMC  
can mix on faster time scales than their collapsed counterparts. \\

\subsection{Reversible MCMC in one-dimensional N-particle systems}
\label{subsec:NPartOneDimReversible}

Although local hard-sphere MCMC was introduced many decades 
ago~\cite{Metropolis1953}, its mixing times were obtained rigorously only in 
recent years, and this only in one spatial dimension. In the discrete case (the 
symmetric simple exclusion process), the especially precise description 
of the convergence process calibrates numerical approaches. \\

\subsubsection{Reversible discrete one-dimensional MCMC}
\label{subsec:OneDDiscreteReversible}

The symmetric simple exclusion process
implements the local hard-sphere Metropolis algorithm in the discrete sample 
space
$\Omega^{\text{SSEP}} = \SET{x_1 < x_2< \TO <  x_N}$ with $x_i \in \SET{1 \TO 
n}$ (periodic boundary conditions understood for positions 
and indices). All legal hard-sphere configurations $c$ 
have the same weight $\pi_c = \pi^*$.
At each time step $t$, a random move $x_i^t \to \xtilde_i = x_i^t \pm 1$ is 
proposed for a (single) randomly sampled particle $i$.
If vertex $\xtilde_i$  is unoccupied, the move is accepted 
($x_i^{t+1} =  \xtilde_i$). Otherwise the particle stays put
($x_i^{t+1} =  x_i^t$).
The process is time-reversible and it thus  trivially satisfies detailed 
balance. The equilibrium flows into a legal configuration $c = (x_1 \TO x_N)$ 
arrive from the configuration $c$ itself and possibly from $2N$ neighboring 
configurations that may however not all be legal~\cite{KapferKrauth2017}:
\begin{equation}
\begin{aligned}
 c^+_i& = (x_1 \TO x_i - 1 \TO x_N)\\
 c^-_i& = (x_1 \TO x_i + 1 \TO x_N).
\end{aligned}
\label{equ:SSEPInFlow}
\end{equation}
The configuration $c_i^+$ contributes flow $\ACAL^+_i = \frac{1}{2N} 
\pi_{c_i^+}$, which is $\frac{1}{2N} \pi^*$ if $c_i^+$  is legal, and zero 
otherwise.
(The factor $\frac{1}{2N}$ accounts for the probability of sampling the index
$i$ and one of the two directions).
Likewise, $c$ contributes flow $\RCAL_i^- = \frac{1}{2N} \pi^*$
through a rejected backward move $ c \to c_i^+$, but only if $c_i^+ $ is 
illegal. An analogous exclusion relation exists for $c_i^-$, 
so that the total flow into configuration $c$ is:
\begin{equation}
\text{Total flow into $c$:}\quad
 \sum_{i=1}^N ( \underbrace{\ACAL_i^+ +  \RCAL_i^-}_{\frac{1}{2N} \pi^*} +  
\underbrace{\ACAL_i^- + \RCAL_i^-}_{ \frac{1}{2N} \pi^*}  ) = \pi^* \equiv 
\pi_c. 
\label{equ:FlowsSSEP}
\end{equation}

The mixing time of the symmetric simple exclusion process from the most 
unfavorable initial distribution $\pit{0}$ (the compact configuration) is known 
to scale as \bigObs{N^3 \log N}, whereas the mixing time from an equilibrium 
configuration scales only as \bigObs{N^3}, that is, as the correlation 
time~\cite{Lacoin2016detailed}. These behaviors are recovered by 
numerical simulation.  \\

\subsubsection{Continuous one-dimensional reversible MCMC}
\label{subsec:ContinuousOneDReversible}

In the one-dimensional continuum, the scaling for the mixing times of 
reversible local hard-sphere MCMC has been obtained for the heat-bath 
algorithm. In this dynamics, at each time step, the position of a randomly 
sampled particle $i$ is updated to a random value in between particles $i-1$ 
and 
$i+1$. The heatbath algorithm mixes in \bigObs{N^3\log{N}} 
moves~\cite{RandallWinklerInterval2005} on an interval with fixed boundary 
conditions. The mixing time for the same model with periodic boundary 
conditions is between \bigObs{N^3} and 
\bigObs{N^3\log{N}}~\cite{RandallWinklerCircle2005}. Numerical computations 
confirm \bigObs{N^3\log{N}} mixing on the continuous interval with periodic 
boundary conditions both for the heatbath and for the Metropolis 
algorithm~\cite{KapferKrauth2017}. \\

\subsection{Non-reversible MCMC in one-dimensional N-particle systems}
\label{subsec:NPartOneDimNonReversible}

Non-reversible MCMC in one-dimensional particle systems has a long history in 
physics and mathematics through the study of the totally symmetric simple 
exclusion process (TASEP)~\cite{ChouTASEP2011}, which is one half of a 
directional lifting of the simple symmetric exclusion process. Even for that 
model, however, 
rigorous mixing-time results are very recent. The \quot{particle-lifted} 
TASEP~\cite{KapferKrauth2017} is a variant of the TASEP and, at the same time, 
a finite-displacement variant of ECMC.  With its chains of particles moving 
forward, it features intriguing mixing behavior. 
This introduces to the problem of 
restarts, a key subject for ECMC. 

All systems considered in this subsection have periodic boundary conditions. 
In the setting where the collapsed algorithm features forward and backward 
moves, one may access the limit of vanishing switching rates, that is, 
retain only one of the equivalent sectors of the lifted transition matrix.  \\

\subsubsection{Discrete non-reversible MCMC in one dimension - TASEP, 
lifted TASEP}
\label{subsec:TASEP}

The TASEP is a displacement lifting of the simple symmetric exclusion process 
(see \subsect{subsec:OneDDiscreteReversible}) with a lifted sample space 
$\Omegahat = \Omega ^{\text{SSEP}} \times \SET{+,-}$. Because of 
the periodic boundary conditions, 
neither the global-balance 
condition nor the irreducibility condition require liftings between 
the \quot{$+$} copy, where particles attempt moves as $x_i \to x_i +1$ 
and the \quot{$-$} copy, where particles attempt moves  as $x_i \to x_i -1$.
It suffices to consider only one 
half of the lifted sample space, say, the \quot{$+$} copy.

The TASEP attempts at each time step to advance a random particle $i$ from its 
current vertex $x_i^{t}$ to $\xtilde = x_i^{t} +1$ (with periodic boundary 
conditions), in other words from $c$ to $c_i^-$.
If the vertex $\xtilde_i$ is  unoccupied (if $c_i^-$ is legal), 
the move is accepted ($x_i^{t+1} = \xtilde_i$), and otherwise the particle 
stays put.

The equilibrium flow into a configuration $c = (x_1 \TO x_N)$ now arrives from 
$c$ itself and possibly from the $N$ neighboring conditions $c_i^+$ of 
\eq{equ:SSEPInFlow}. The configuration $c_i^+$ contributes accept-flow 
$\ACAL_i^+ = \frac{1}{N} \pi_{c_i^+}$ which is $\frac{1}{N} \pi^*$ if $c_i^+$ 
is legal, and zero otherwise (compare with 
\subsect{subsec:OneDDiscreteReversible}). Now, $c$ itself contributes 
reject-flow of particle $i-1$, namely $\RCAL_{i-1}^+$, 
which is $\frac{1}{N} \pi^*$ if $c_i^+ $ is illegal and zero 
otherwise. The total flow into configuration $c$ (with periodic boundary 
conditions understood) is therefore~\cite{KapferKrauth2017}: 
\begin{equation}
\text{Total flow into $c$:}\quad
 \sum_{i=1}^N ( \underbrace{\ACAL_i^+ +  
\RCAL_{i-1}^+}_{\frac{1}{N} \pi^*} ) = \pi^* \equiv \pi_c.
\label{equ:FlowTASEP}
\end{equation}

The TASEP is proven to mix in \bigObs{N^{5/2}} (see~\cite{BaikLiu2016}). The 
equilibrium correlation time (the inverse gap of the transition matrix that, in 
this case, can be diagonalized) has the same \bigObs{N^{5/2}} 
scaling~\cite{Dhar1987,GwaSpohnPRL1992}. Numerical 
simulations (with an operational definition of mixing 
times~\cite[Supplementary Item S2]{KapferKrauth2017}) recover this behavior.

In \eq{equ:FlowTASEP}, the rejected moves of particle $i-1$ compensate the 
accepted moves of particle $i$, because in the TASEP they are both sampled with 
the same $\frac{1}{N}$ probability. For this reason, a naive sequential TASEP, 
where particles are simply chosen one after the other, is incorrect. The flow 
into a given lifted configuration $(c,i)$ is given by terms analogous to 
$\ACAL_i^ +$ and to $\RCAL_i^+$ (see \eq{equ:FlowTASEP}, but without the 
factors $\tfrac 1N$). They do not add up to a constant $\pi^*$. An improved 
lifting scheme however remedies the situation. The particle-lifted 
TASEP~\cite{KapferKrauth2017} concerns a lifted sample space $\Omegahat = 
\Omega ^{\text{SSEP}} \times \SET{1^+ \TO N^+; 1^- \TO N^-}$. Again, the 
periodic boundary conditions allow the \quot{$+$} and \quot{$-$} sectors of 
lifting variables to separate, and only one of these sectors (and of the 
corresponding transition matrix) must be considered, say, the \quot{$+$} 
sector.  In the particle-lifted TASEP,  the lifted configuration $(c, i^+)$ 
attempts to move to $(c^-_i,i^+)$ (by advancing $i$ by one vertex). If $c^-_i$ 
is illegal, the particle lifting $ (c, i^+) \TO (c, (i+1)^+)$ takes place. The 
flows into a configuration $\chat = (c, i^+)$ is contributed to by two possible 
configurations, namely by $(c, (i-1)^+$ and by the configuration $(c^+_i,i)$ of 
\eq{equ:SSEPInFlow}. Global balance can be established analogously to 
\subsect{subsec:TASEP}, with the uniform stationary distribution $c^*$. The 
particle-lifted TASEP is deterministic, and is not irreducible (a compact state 
of $N$ particles evolves at best into a compact state of $N-1$ plus one 
detached particle). Randomness can be introduced through 
restarts~\cite{KapferKrauth2017}. \\

\subsubsection{Continuous non-reversible MCMC in one dimension: Forward, Lifted 
Forward}
\label{subsec:NonRevSeqForward}

The discrete MCMC algorithms of \subsect{subsec:TASEP} can all be generalized 
to the continuous case with $\Omega^{\text{SSEP}} $ replaced by the 
$N$-dimensional hypercube  with $0< x_i < x_{i+1} < x_N <L $ (with periodic 
boundary conditions in $L$ implied). The moves $x_i \to x_i \pm 1$ of the 
discrete Markov chains can now be generalized to $x_i \to x_i \pm \epsilon$, 
where $\epsilon > 0 $ is taken from a certain probability distribution and the 
non-hopping condition  $x_i < x_{i+1}$ (corrected for periodic boundary 
conditions in $N$) is enforced. 

With the non-hopping condition, the MCMC algorithms  for spheres of diameter 
$\sigma$ on a ring of length $L$ are equivalent to those for zero-diameter 
point particles in a ring of length $\Lfree =   L -  N \sigma$, for which the  
non-hopping condition becomes the only test of legality of a configuration. The 
global-balance condition can be checked for every value of $\epsilon$ 
separately, with configurations $c_i^{+\epsilon}$ and $c_i^{-\epsilon}$ 
generalizing the $2N$ neighboring configurations $c_i^{+}$ and $c_i^{-}$ of 
\eq{equ:SSEPInFlow}. The value of $\epsilon$ may serve as an additional lifting 
variable. This has however not been studied.

The forward Metropolis algorithm, where a move by $\epsilon>0$ is proposed to a 
random sphere $i$, is the continuous-space version of the TASEP, and the 
particle-lifted forward Metropolis algorithm generalizes the particle-lifted 
TASEP. As this algorithm now has a random element $\epsilon$, it is generically 
irreducible and aperiodic, and its mixing time is 
numerically found to scale as $\tmix = \bigObs{N^{5/2}}$ 
(see~\cite{KapferKrauth2017}). With a suitable restart strategy, this mixing 
time can be decreased to $\bigObs{^2 \log N}$ scaling. The sequential version 
of the lifted forward Monte Carlo algorithm is again incorrect, although the 
local relabeling strategy outlined in \subsect{subsec:ECMCRelabeling} can be 
applied. Extensions of the hard-sphere case to nearest-neighbor, monotonous 
potentials with a no-hopping conditions were discussed~\cite[Supp. Item 
5]{KapferKrauth2017}. \\

\subsection{ECMC in one-dimensional N-particle systems}
\label{subsec:NPartOneDimLifted}

For one-dimensional $N$-particle hard-sphere systems, ECMC is the 
infinitesimal-displacement limit ($\epsilon \to 0$)  of the particle-lifted 
forward Metropolis algorithm (see \subsect{subsec:NonRevSeqForward}). Its 
validity follows, in addition to this limiting procedure and to an explicit 
check of global balance, through the equivalence of this specific MCMC 
algorithm with molecular dynamics with an initial condition where one sphere 
has velocity $+1$ and all other spheres are stationary (velocity $v_i=0$). 
In the $\epsilon \to 0$ limit, the details of the probability distribution of 
$\epsilon$ disappear, and only its mean value remains relevant. The randomness 
in $\epsilon$ that ensures the irreducibility of the forward Metropolis 
algorithm disappears. 
ECMC thus requires restarts after a given chain length $\ell = \sum \epsilon_t$
in order to achieve irreducibility. These restarts replace the re-sampling 
transition  matrix $\Pres$ of \subsect{subsec:BoundedOneDSingle}. For specific 
chain-length distributions (probability distributions for  $\ell$), the ECMC 
relaxation can be described 
exactly. Non-trivial relabelings and factor-field variants again illustrate 
that ECMC is a family of algorithms with different mixing-time scalings rather 
than a fixed dynamics. \\

\subsubsection{Standard ECMC, coupon-collector problem}
\label{subsec:Coupon}

Hard-sphere ECMC in one dimension (with periodic boundary conditions) consists 
in the limit of the forward Metropolis algorithm with restarts, where the step 
size $\epsilon_t$ at time $t$ approaches zero, and the number $l$ of steps in 
the event chain (between restarts) approaches infinity such that their product 
the  total chain length $\ell = \sum_{t=0}^{l-1} \epsilon_t$, remains finite. 
For hard spheres, ECMC is devoid of random elements except of the chain length 
$\ell$ (which may be a random variable) and the sphere which initiates each 
event chain.

In the reduced-circle representation of length $\Lfree$, with spheres of zero 
diameter (see \subsect{subsec:NonRevSeqForward}), one event chain of length 
$\ell$ effectively advances the sphere that initiates the event chain  by a 
distance $\ell$. For $\ell = \ran[\const, \const + \Lfree]$, the initial sphere 
of the event chain is placed (up to a relabeling) at a random position on the 
ring. EMCMC obtains a perfect sample when each sphere has once initiated an 
event chain. The solution of the mathematical coupon-collector problem shows 
that it takes \bigObs{N \log N} random samples of $N$ elements to touch each of 
them at least once. In consequence,  the number of event chains required for 
mixing scales as $\sim N \log N$, with a total event count of \bigObs{N^2 \log 
N} (see \cite{Lei2018_OneD,Lei2018PHD}. This number of events represents the 
computational effort more faithfully than the number of time steps, which 
diverges (by construction) for each finite-length event chain. \\

\subsubsection{ECMC with local relabeling}
\label{subsec:ECMCRelabeling}

Pair-factor ECMC may be modified by having at each event the two particles of 
the vetoing factor not only switch their lifting status (active / target), but 
also their identities (particle indices). This relabeling is without incidence 
on the sequence of positions during a given event chain, throughout which the 
initial sphere remains active. The relabeling at each event is local. It 
suffices to render sequential ECMC correct~\cite{Lei2018_OneD}. (The same 
relabeling also allows the sequential forward MCMC to satisfy global balance.) 
With the initial sphere of each event-chain  taken sequentially rather than 
randomly, the coupon-collector-related logarithmic slowdown is avoided so that 
the choice of random chain lengths $\ell = \ran[\const, \const + 
L_{\text{free}}]$, the only remaining random element in the algorithm, produces 
a perfect sample after $N$ chains and \bigObs{N^2} events rather than the 
\bigObs{N^2 \log N} events of regular ECMC. ECMC with local relabeling again 
illustrates that reducing the randomness of moves can speed up mixing, in other 
words, bring about faster overall randomization. In higher-dimensional 
hard-sphere systems, the relabeling appears to speed up the approach to 
equilibrium by only a constant factor~\cite{Lei2018PHD}. \\

\subsubsection{Factor-field ECMC in  one dimension}
\label{subsec:FactorFieldsOneD}

The factorized Metropolis filter separates the total potential into independent 
terms (the factors). Large variations in one factor potential then remain 
uncompensated by those in other factors,  as would be the case for the total 
potential (for energy-based filters). As a consequence, ECMC may possess event 
rates that are too high for efficient mixing. This potential shortcoming of 
standard ECMC was remarked at low temperature in one-dimensional $N$-particle 
systems~\cite{Hu2018} (see also~\cite[Fig. 2]{Lei2018_OneD}). It is overcome in 
one-dimensional particle systems through the addition of invariants and their 
subsequent breakup into factor fields. Such terms may decrease the event rate. 
Moreover, they can profoundly modify the characteristics of the event-chain 
dynamics and  decrease mixing-time and correlation-time exponents.

For $N$ particles on a one-dimensional interval of length $L$, the quantity 
$\sum_i(x_{i+1} - x_{i})$ (with periodic boundary conditions in $L$ and in $N$) 
is an invariant. A linear function $f(x) = ax$ can transform it into a sum of 
factor fields:
\begin{equation}
 \underbrace{f\glc \sum_i \glb x_{i+1} - x_i \grb \grc}_{\text{invariant}}  = 
\sum_i \underbrace{f\glb x_{i+1} - x_i \grb}_{\text{factor field}}
\label{equ:InvariantFactorField} 
\end{equation}
The factor field of \eq{equ:InvariantFactorField} may be added and its linear 
parameter $a$ adjusted to any pair-factor potential. Attractive factor fields 
may thus be added to hard-sphere factors or to Lennard-Jones 
factors~\cite{Lei2019}. With the linear factor adjusted to compensate the 
virial pressure, \bigObs{N^{3/2}} autocorrelation times (rather than 
\bigObs{N^2} without factor fields) and \bigObs{N^2} mixing times (rather than 
\bigObs{N^2 \log N} are found. One particular feature of event-chain dynamics 
at the optimal value of the factor field is that the chains have zero linear 
drift, which itself is a measure of the virial 
pressure~\cite{Lei2019,Michel2014JCP}. \\

\section{Statistical-mechanics models in more than one dimension}
\label{sec:StatMechgreater}

The present section reviews statistical-physics models in more than one 
dimensions where, in ECMC, several factors are present at any time. The 
vetoing factor is chosen among them. \subsect{subsec:TwoDimHard} reviews the 
two-dimensional hard-disk model, the first application of  
ECMC~\cite{Bernard2009,Bernard2011,Bernard2011PHD}. The long-range physics of 
this model is intricate, with phonons coexisting with two types of topological 
excitations (dislocations and disclinations). Hard-disk ECMC is contrasted with 
related algorithms that have been applied to this system. 
\subsect{subsec:Harmonic} briefly reviews ECMC for the harmonic model of a 
solid (with a fixed list of neighbors for each particle) which also serves as 
the fundamental low-temperature theory for continuous-spin systems, as the XY 
or the Heisenberg model. In these spin models, the behavior of ECMC has been 
thoroughly analyzed, and the interplay between spin waves and topological 
excitations for local MCMC algorithms including ECMC, as reviewed in 
\subsect{subsec:XY_Heisenberg}, is better understood than for particle systems.
\\

\subsection{Two-dimensional hard disks}
\label{subsec:TwoDimHard} 

ECMC is conceptually simpler for hard spheres than for general interactions, as 
the potential switching between zero and infinity is deterministic rather than 
stochastic (see \subsect{subsec:StochasticPotentialSwitching}). Most studies 
have concentrated on the two-dimensional case, the hard-disk model, where 
fundamental results were obtained using ECMC~\cite{Bernard2011,Engel2013}. (The 
original Fortran90 implementation used is publicly available~\cite{Li2020}.) 
Hard-disk ECMC is found to be very fast, but elementary questions as for 
example the scaling of correlation times with system size, in the hexatic and 
solid phases, are still without even an empirical answer. The simplifications 
related to the existence of a constraint graph of possible liftings are at the 
heart of a recent implementation of multithreaded ECMC for hard 
disks~\cite{Li2020}. \\

\subsubsection{Characterization of hard-sphere ECMC}
\label{subsec:TwoDimensionPhysical}

Pair-factor ECMC for two-dimensional hard disks~\cite{Bernard2009} consists in 
freely moving an active sphere in continuous time $t$ along a fixed direction 
up to  a lifting event. This event corresponds to contact with  a target 
sphere, which then exchanges its lifting status with the active one. Two 
versions of hard-sphere ECMC are compatible with the conservation of 
phase-space 
volume: \quot{Straight} ECMC, where the out-state displacement is along the 
same directions as the in-state displacement, reducing the effective spatial 
dimension to one. Between restarts, when all lifting variables are reset, 
straight ECMC is in higher dimensions no different from the one-dimensional 
case treated in \sect{subsec:NPartOneDimLifted}. For straight ECMC, the lifted 
sample space is for example $\Omegahat = \Omega \times \SET{1^{\pm x} \TO 
N^{\pm 
x}; 1^{\pm y} \TO N^{\pm y}}$, where $2N$-dimensional collapsed sample space 
$\Omega $ describes the particle positions. Periodic boundary conditions are 
assumed, and the \quot{$+x$} and \quot{$+y$} copies separate from the motions 
in 
the negative directions, which can be neglected.  The correctness of straight 
ECMC also follows from the continuum limit of a hard-disk lattice version as 
well as from a mapping onto molecular dynamics with two-dimensional sphere 
positions but with one-dimensional sphere velocities (see for 
example~\cite[Lemma 1]{Li2020}). Alternatively, \quot{reflected} ECMC uses as 
an 
out-state displacement direction the one reflected with respect to the line 
(hyperplane) of incidence (see~\cite[Fig. 2.8]{Bernard2011PHD}. Straight ECMC 
is 
generally faster~\cite[Fig. 6]{Bernard2009}. A variant of straight ECMC with a 
larger direction space $\DCAL$ containing many angles was found to bring no 
improvements for hard disks~\cite{Weigel2018}, while it has drastic effects for 
hard-sphere objects with internal degrees of 
freedom~\cite{Qin2020fastsequential}. 

In hard-sphere systems, the computation of the pressure is notoriously 
complicated for generic MCMC, which does not allow one to compute the virial 
from a finite number of equiprobable configurations. Instead, the virial is 
extracted from a high-precision extrapolation of the pair-correlation function 
to contact~\cite[Sect. 3.3.4]{Bernard2011PHD}. Continuous-time ECMC obtains an 
infinite number of highly correlated samples between any two lifting events, 
and it allows for an improved estimator~\cite[Eq. 20]{Michel2014JCP}) for the 
virial pressure from the expectations of basic geometrical properties of the 
ECMC trajectories. The estimator was tested in two-dimensional and 
three-dimensional hard spheres 
(see~\cite{Michel2014JCP,Engel2013,IsobeKrauth2015}. It generalizes from hard 
spheres to arbitrary potentials.

Between restarts of hard-disk ECMC, any disk can lift with no more than three 
other disks, if they are monodisperse (see~\cite{KapferPolytope2013}). This 
generalizes from the one-dimensional case, where the order between spheres 
remains unchanged so that a sphere lifts  with only a single other sphere that 
it cannot hop over. A constraint graph encodes these relations. It has at most 
three outgoing arrows for every disk, corresponding to the liftings it can 
provoke (see~\cite{KapferPolytope2013}). The optimal (sparsest) constraint 
graph is locally planar~\cite[Sect. 3.1]{Li2020}. The constraint-graph 
formulation of hard-sphere ECMC exposes its close connection with the 
harmonic-model ECMC (see \subsect{subsec:Harmonic}), where the neighbor 
relations are likewise fixed. 

In two-dimensional hard-disk MCMC algorithm, the total variation distance or 
spectral gaps cannot be estimated or evaluated, and theoretical bounds cannot 
be used~\cite{DiaconisLebeauMichel2011}. Rigorous mixing-time scaling exponents 
are 
available for low density, but only for a non-local version of the Metropolis 
algorithm~\cite{Kannanrapidmixing2003}. The analysis of all recent computations 
builds on the hypothesis that for two-dimensional disks the autocorrelation 
function of the  global orientational order parameter is the slowest relaxation 
process in this system. The hypothesis has proven robust. Practical 
computations generally adopt a square box with periodic boundary conditions, 
where the expectation of the above autocorrelation function vanishes because of 
symmetry, simplifying the interpretation of time 
series~\cite{Bernard2009,Bernard2011}. \\

\subsubsection{Hard-disk ECMC and other algorithms}
\label{subsec:TwoDimensionComparison}

In ECMC, the continuous-time limit is usually required in order to break ties 
between candidate event times of different factors. For two particles with pair 
interactions in a box with periodic boundary conditions,  and in particular for 
two hard spheres, ECMC satisfies global balance with finite displacements 
$\DCAL = \SET{\delta_x, \delta_y}$ (that would have to vary in order to ensure 
aperiodicity). Because of translational invariance, a given event chain, say, 
with a fixed displacement $\delta_x$ then maps to the transition matrix 
$\Ptrans$ on a path graph (see \subsect{subsec:Diaconis}), with the 
displacement 
of one hard disk corresponding to the \quot{$+1$} sector, and the displacement 
of the other corresponding ot the \quot{$-1$} sector. 

For any $N$ and in any spatial dimension $d$, hard-sphere ECMC is equivalent to 
modified molecular dynamics where, for one event chain, the positions are 
$d$-dimensional but the velocities are one-dimensional. This corresponds to 
hard spheres on one-dimensional constraining \quot{rails} that remain fixed in 
between restart. The ECMC events correspond to molecular-dynamics collisions, 
which conserve energy and momenta for an configuration with only a single 
non-zero velocity (the one of active disks). 

A discrete-time precursor algorithm of ECMC~\cite[Sect. 5]{Jaster1999} chooses 
for each move a random initial disk, a random direction of displacement (such 
that a direction and its inverse are sampled with the same probability), and a 
total number $n_c$ of disks to be displaced. A chain move is then constructed 
(not unlike ECMC) by displacing $n_c-1$ disks (starting with the initial one) 
along the direction of displacement until they  hit their successor disks. The 
$n_c$th (final) disk is placed randomly between its initial position and its 
(hypothetical) successor disk. In order to satisfy detailed balance, this 
algorithm requires a Metropolis rejection step for the entire chain by the 
ratio 
of the intervals available for the first and for the final disks. One move of 
the precursor algorithm resembles the continuous-time ECMC evolution between 
restarts. However, it cannot be interpreted in terms of a 
continuous-time 
Markov chain, and it requires 
rejections and must allow for chains in a given direction and its inverse 
direction with equal probability. The convergence properties of this algorithm 
have not been analyzed. 

The hard-disk model was also successfully simulated on graphics cards (GPUs) 
with a massively parallel implementation of the Metropolis algorithm using a 
four-color checkerboard scheme. In this scheme, a cell of any one color touches 
only differently colored 
cells~\cite{AndersonGPU2013,Engel2013}). At any  time, disks in cells 
of only one color can thus be updated in parallel. Irreducibility is assured by 
frequently translating the four-color checkerboard by a random vector. This 
highly parallel algorithm overcomes its considerable speed handicap (roughly 
two orders of magnitude in single-processor mode), through massive 
parallelization. Massively parallel MCMC has been tested to high precision 
against ECMC and against event-driven molecular dynamics for a number of 
physical quantities, as the pressure (that is, the equation of state) and the 
orientational and positional order parameters~\cite{Engel2013}.  \\

\subsubsection{Parallel hard-disk ECMC}
\label{subsec:TwoDimensionParallel}

Hard-disk ECMC and event-driven molecular dynamics both identify the earliest 
one of a number of candidate events, the one that will be realized as an event 
and drive the dynamics of the system. The event then generates new candidate 
events, while some of the old ones continue to exist and yet others disappear. 
Modern event-driven molecular dynamics codes are optimized for 
the management and the update of a very large number of such candidate events, 
typically organized in a heap~\cite{Rapaport1980,Isobe2016}. In event-driven 
molecular dynamics, an extensive number $\propto N$ of candidate events are 
present at any given moment. The possible scheduling conflicts among this large 
number of candidate events has long stymied attempts to  parallelize the 
algorithm, that is, to handle several events independently from each 
other~\cite{Lubachevsky1992,Lubachevsky1993, 
Greenberg1996,Krantz1996,Marin1997}. 

Domain-decomposition strategies for the candidate events in molecular dynamics 
present many problems of their own~\cite{Miller2004}. Within 
ECMC~\cite{KapferPolytope2013}, domain decomposition leads to residual 
interactions that destroy the global translational symmetry, so that the very 
convenient separation into non-communicating sectors of lifting variables 
breaks down, and some of the ECMC efficiency is lost.

In ECMC, one may freely choose the number of active particles (those with 
non-zero velocities) and keep this number fixed throughout a simulation. If 
this number is  $\lesssim \sqrt{N}$, the mathematical birthday problem shows 
that the candidate events are usually disjoint for any two active particles, 
cutting down on the degree of interference between different active particles.  
Using a framework of local times, this has allowed to prove for 
that scheduling conflict hard-sphere systems appear with finite probability in 
the $N \to \infty$ limit for finite run times~\cite{Li2020}. \\

\subsection{Harmonic model}
\label{subsec:Harmonic} 

The harmonic model~\cite{Wegner1967} describes spin-wave excitations for $N$ 
spins on a lattice, with a total potential $U = \half \sum_{\neigh{i}{j}} 
(\phi_i - \phi_j)^2$ between neighbors $\neigh{i}{j}$, with non-periodic angles 
$\phi_i \in \RR$, which approximate the small elongations $|\phi - \phi_j| \ll 
1$ in the XY model with its total potential $U = - \sum_{\neigh{i}{j}} 
\cosb{\phi_i - \phi_j}$. Spin waves are the dominant excitations in the XY 
model at low temperatures, notably in two 
dimensions~\cite{FroehlichSpencer1981}. The harmonic model also provides the 
quintessence of phonon excitations in particle systems, for small displacements 
from  perfect lattice positions. In the harmonic particle model, each particle 
interacts harmonically with a fixed set of neighbors so that disclinations and 
stacking faults, and other excitations, cannot develop. 

Besides its role of isolating phonon (resp. 
spin-wave) excitations of many systems, the harmonic model is of importance for 
hard-sphere ECMC whose sequence of constraint 
graphs~\cite{KapferPolytope2013,Li2020}, between resamplings,
effectively defines a sequence of models with fixed neighborhood. \\

\subsubsection{Physics of the harmonic model}
\label{subsec:HarmonicPhysics}

The harmonic model is exactly  solved~\cite{Wegner1967}. It has a single phase 
for all finite temperatures, but the nature of this phase depends on 
dimension $d$. The differences in angle for two spins at 
positions distant by $\rvec$ for the harmonic spin model (and similarly the 
difference in elongations with respect to the lattice positions for the 
harmonic 
particle model) in equilibrium are given by:
\begin{equation}
\underbrace{ \mean{(\Delta \phi) ^2}}_{\text{spin harm.}} \sim
\underbrace{ \mean{(\Delta \xvec)^2}}_{\text{part. harm.}}  \propto 
\begin{cases}
r & \text{for dimension $d=1$}\\
\log r   & \text{for $d=2$}\\
\const   & \text{for $d \geq 3$}.
\end{cases}
\label{equ:HarmonicFluctuations}
\end{equation}
For a system of size $L$, the mismatch of two typical spins or particles is 
thus 
of the order of \bigObs{\sqrt{L}} in one spatial dimension, it grows as the 
square root of the  logarithm of $L$ in two dimensions, and it remains constant 
in three dimensions and higher. The harmonic particle model, in two dimensions, 
features long-range orientational order but only power-law decay of positional 
order~\cite{Mermin1968}. Only in more than two dimensions does it have 
long-range orientational and positional order.  \\

\subsubsection{ECMC algorithm for the harmonic model}
\label{subsec:HarmonicECMC}

The ECMC algorithm for the harmonic spin model performs in displacement and 
particle lifting. Each lifted configuration is described by $\SET{\phi_1 \TO 
\phi_N} \times \SET{i, \DCAL}$, where $i$ describes the active spin and $\DCAL 
\in \SET{+,-}$ the direction of changing the angular variable $\phi$ (which is 
between $-\infty$ and $\infty$). Because of the invariance of the pair 
potentials with respect to the absolute spin angles, no liftings between $\DCAL 
= +$ and $\DCAL = -$ are required, and the lifted sample space $\Omegahat$ 
again
splits into two equivalent copies. It suffices to retain the \quot{$+$} 
direction.o te The harmonic model essentially differs from the spin model 
in that the 
$2 \pi$ periodicity is lost. ECMC monotonically increases each of the $\phi_i$, 
at a difference of Metropolis MCMC, which must allow for changes of the angles 
in both directions. In event-driven harmonic-model ECMC, for a finite number of 
neighbors per spin, the displacement per event of each active spin is finite. 
It 
can be expected that the total displacement of each spin or particle 
corresponding to the $\Delta \phi$ in \eq{equ:HarmonicFluctuations} 
decorrelates 
the system. It thus follows that each particle or spin moves between 
independent 
samples \bigObs{\sqrt{L}} times in one dimension, \bigObs{\sqrt{\log{L}}} times 
in two dimensions, and a constant number of times in more than two dimensions, 
whereas reversible local MCMC requires $\propto L^2$ displacements per spin (or 
particle) for all dimensions. This translates into correlation times of 
\bigObs{N^{3/2}}, \bigObs{N \sqrt{\loga{N}}} and \bigObs{N} single events, in 
dimensions $d=1,2,3$. Numerical simulations are in excellent agreement with 
these expectations~\cite[Fig. 7]{Lei2018_TOP}. \\

\subsection{ECMC for continuous spin models}
\label{subsec:XY_Heisenberg}

ECMC readily applies to XY 
and Heisenberg-type spin models because their spin--spin pair potentials
on neighboring sites $i$ and $j$  write as $U(\phi_i - \phi_j)$ with continuous 
spin angles $\phi_i$ and $\phi_j$. The invariance of the pair potentials with 
respect to the absolute spin angles then simplifies ECMC 
in the same way that translational invariance does for particles. For 
spin models, the algorithm rotates the spin $i$ in a given sense until the 
factorized Metropolis algorithm calls for a veto by a neighbor $j$ of $i$. The 
spin $i$ then stops and $j$ starts to rotate in the same sense, with the 
cumulative rotation corresponding to the MCMC time. 

In the XY model, the two-dimensional fixed-length continuous spins 
live on a $d$-dimensional spatial lattice.
The pair 
potential $U_{ij}= -\cosb{\phi_i - \phi_j}$ favors the alignment of 
neighboring spins $i$ and $j$, and in ECMC, the activity (that is, the 
rotation) passes from one spin to one of its neighbors. ECMC needs no 
restarts, and its 
sequence of active spins realizes, in the two-dimensional lattice,  an 
anomalous 
diffusion process~\cite{KimuraHiguchi2017}. ECMC on the Heisenberg model, 
where spins are three-dimensional, can be reduced to this case, with restarts, 
through rotations in two-dimensional sub-planes in spin-space.  In both 
models, ECMC lowers the relaxation rates with 
respect to local reversible MCMC. Although more efficient 
algorithms are available~\cite{HasenbuschXY2005,Hasenbusch_Clock2019,Xu2019}
for these particular models, the comparison of ECMC with reversible MCMC may 
illustrate what can be achieved through non-reversible MCMC in real-world 
applications where, as 
discussed in \sect{sec:Introduction}, customized \emph{a priori} move sets 
remain unavailable. 

In the two-dimensional XY model, ECMC autocorrelation functions can be fully 
explained in terms of spin waves and topological excitations. The 
two-dimensional hard-disk system similarly features phonons and two types of 
topological excitations, but the precise relaxation dynamics of MCMC algorithms 
has yet to be clarified (for a synopsis, see~\cite[Chap. 1]{Bernard2011PHD}).
\\

\subsubsection{Spin waves and topological excitations in the two-dimensional XY 
model}
\label{subsec:SpinWavesTopologicalXY}

Although it possesses no long-range spin order at finite temperatures, the 
two-dimensional XY model famously undergoes a phase 
transition~\cite{KosterlitzThouless1973,HasenbuschXY2005} between a 
low-temperature 
phase rigorously described by spin waves with bound vortex--antivortex 
pairs~\cite{Wegner1967,FroehlichSpencer1981}, and a high-temperature phase 
where these topological excitations are free.

In the low-temperature phase, vortices pair up with antivortices. 
The maximum $ d_{\max}$ of all vortex--antivortex pair separations in a system 
of size $L \times L$ follows a Fréchet distribution~\cite[]{Lei2018_TOP}, 
\begin{equation}
\probb{d_{\max}} = \frac{\alpha}{s} \glb \frac{d_{\max}}{s} \grb ^{-1-\alpha}
\expc{- \glb \frac{d_{\max}}{s} \grb^{- \alpha} },
\label{equ:Frechet}
\end{equation}
with a size-dependent scale $s = L^{2/\alpha} s_0$ and a size-independent 
exponent $\alpha$ that depends on the inverse temperature $\beta$. Throughout 
the low-temperature phase, $d_{\max} \sim L^{2/\alpha}$, with $\alpha > 2$ 
increases slower than the system size $L$ for non-zero temperatures. Only in 
the zero-temperature limit does $ d_{\max}$ remain constant as the system size 
increases ($\alpha \to \infty$ for $T \to 0$). It is extensive as the 
transition point is approached ($d_{\max} \sim L$, that is, $\alpha \to 2$ for 
$T \to T_c$). \\

\subsubsection{Relaxation time scales in spin models}
\label{subsec:SpinWavesDoubleTimeScale}

In spin models, local MCMC simulations must relax spin waves and topological 
excitations. In the two-dimensional XY model  of length $L$, reversible 
energy-based MCMC can be expected to diffusively relax spin waves on a 
\bigObs{L^2} time scale at all temperatures~\cite{Lei2018_TOP} and topological 
excitations on a \bigObs{d_{\max}^2} time scale, so that the overall 
correlation 
time scales as \bigObs{L^2} from both contributions. ECMC relaxes spin waves on 
a much faster time scale that, from the analogy with the harmonic model, can be 
thought to be \bigObs{\log L}, leading to very fast (system-size-independent) 
initial decay of spin--spin autocorrelations. The diffusive relaxation of 
topological excitations by ECMC on the $\bigObs{d_{\max}} = 
\bigObs{L^{4/\alpha}}$ is dominant at large times throughout the 
low-temperature 
phase, with the two contributions leading to a two-timescale decay of spin 
autocorrelation functions~\cite{MichelMayerKrauth2015,Nishikawa2015}. The above 
arguments are borne 
out by numerical computations~\cite{Lei2018_TOP} with the $\alpha$ of 
\eq{equ:Frechet} as the single nontrivial parameter. \\

\section{ECMC and molecular simulation}
\label{sec:MolecularSimulation}

In molecular simulation of classical atomic models with explicit solvents, 
sampling plays an all-important role~\cite{Sugita1999}, in addition to the 
study 
of explicit time dependence. However, the complexity of total potentials only 
allows for unbiased local moves, and intricate \emph{a priori} choices or 
cluster MCMC algorithms~\cite{Dress1995,LiuLuijten2004} have not met with 
success. The Metropolis or the heatbath algorithms are energy-based: they 
require the evaluation of the total potential. In the Coulomb case, this comes 
with prohibitive time consumption for MCMC. In the field,  molecular dynamics 
codes are therefore used exclusively. Great effort has been directed towards 
the 
computations of the forces (the derivatives of the total potential), which 
remains onerous and fraught with approximations. Molecular dynamics invariably 
implements the Newtonian relaxation dynamics. Several molecular-dynamics codes, 
developed over the last three decades (e.g.~\cite{Plimpton1995}), have found 
score of real-world applications from biology to medicine, physics, and 
chemistry. Molecular 
dynamics is thus highly successful, but it leaves little algorithmic freedom.

ECMC, while still in its infancy, may represent a new starting point for the 
use 
of  MCMC for real-world applications in molecular simulation. First, to sample 
the Boltzmann 
distribution $\pi \propto \expb{-\beta U}$ without approximations, the 
potential 
$U$ (which may contain a Coulomb component) need not be evaluated. Second, in 
the field where, within MCMC, complex \emph{a priori} choices for moves 
continue 
to remain out of reach, the tools of factorization, lifting, and thinning may 
well leverage the development of powerful algorithms. 
\subsect{subsec:MolecularSimulationTheor}, reviews the theoretical aspects of 
ECMC in the context of molecular simulation while \subsect{subsec:Jellyfysh} 
provides an overview of the \quot{JeLLyFysh} Python application, \footnote{The 
name echoes creatures which, like objects studied in molecular simulation, 
mostly contain water, with some other elements, including proteins.} an 
open-source ECMC implementation for molecular simulation. \\

\subsection{Theoretical aspects of ECMC for all-atom models}
\label{subsec:MolecularSimulationTheor}

Three main theoretical challenges stand out in the application of ECMC to 
molecular simulation. First is the choice of factors, in particular of the 
Coulomb factors. The 
different factors are independent and do not compensate each other. Different
types of factorizations 
(especially for the Coulomb potential) therefore impact the total event rate
and the MCMC trajectories
(see \subsect{subsec:ECMCCoulombDipoleFactors}). The second 
challenge for ECMC, reviewed in \subsect{subsec:ECMCCoulomb}, is the handling 
of 
the Coulomb potential. The statistical independence of factors obviates the 
need 
for computing  the total potential. However, the factor event rates (the 
derivatives with respect to the active-particle displacement of the factor 
potentials) must be provided. One such factor may consist in a single pair of 
charges together with all their periodic image charges 
(\quot{merged-image} Coulomb factor) or else in one pair of specific periodic 
image charges (\quot{separate-image} factor). Both choices are practical, with 
the latter leading to an infinite number of factors already for a two-charge 
system~\cite{KapferKrauth2016,Faulkner2018}. For both the merged-image and the 
separate-image formulation, a single 
Coulomb factor may collect potentials between a number of atoms, for example 
all 
those belonging to a pair of neutral molecules.  Such a \quot{dipole-factor} 
choice corresponds at large distances to the interaction of an atom with a 
dipole, or even to the interaction of two dipoles, leading to reduced event 
rates. Finally, the choice of lifting variables (in other words of the set 
$\LCAL$ in $\Omegahat = \Omega \times \LCAL$), as well as lifting schemes (how 
to select the elements of $\LCAL$) appear as main fixtures in a largely 
uncharted territory. As reviewed in \subsect{subsec:ECMCLiftingSchemes}, an 
ECMC 
event of the factor that triggers the veto is analogous to a collision with an 
in-state and an out-state. The out-state is not fixed by a physical dynamics 
and 
must only satisfy the global-balance condition, allowing for the setup of 
in-equivalent lifting schemes for larger factors. The influence of the choice 
of 
lifting variables (for example, the choice of directions, individual moves, 
\emph{etc}.)  on the mixing properties remains poorly understood (see 
\subsect{subsec:ECMCLiftingSchemes}). \\

\subsubsection{Factors, Coulomb factors}
\label{subsec:ECMCCoulombDipoleFactors}

The Coulomb problem illustrates the algorithmic inequivalence of different 
factorizations. With Coulomb pair factors between each individual pair of 
charges distant by $r$ (both for merged-image or separate-image formulations) 
the pair-factor potential is $\sim 1/r$, and the pair-factor event rate $\sim 1 
/ r^2$. In a three-dimensional box of side $L$, at fixed density, so that  $N 
\propto L^3$, a typical event rate between two charges is $\sim 1/L^2$ and the 
total event rate for Coulomb pair factors is $\bigObs{N/L^2} = 
\bigObs{N^{1/3}}$\cite{KapferKrauth2016} (see also \cite[eq. 
(89)]{Faulkner2018}). With this rate, one event corresponds to a single 
particle 
moving forward by $\bigObs{1/N^{1/3}}$. To advance $N$ charges by a constant 
displacement (for example $1$ \AA), Coulomb pair factors require a computing 
effort of $\bigObs{N^{4/3}}$. This is clearly borne out by numerical 
simulations 
(see~\cite{KapferKrauth2016} for details).

For a dipole factor made up of all the atoms in two charge-neutral molecules 
with finite dipole moments, the active particle in one of the molecules 
effectively interacts with a dipole in the other molecule. At a distance $r$ 
between molecules, this gives a factor potential $\sim 1/ r^2$ and a factor 
event rate $\sim 1/r^3$. Integrated over three-dimensional space in a box of 
sides $L$, the total event rate scales as $\bigObs{\log N}$. Advancing $N$ 
charges by a constant distance requires in this case \bigObs{N \log N} events, 
as was observed in $N$-body simulations~\cite[Fig. 13]{Faulkner2018}.

In real-world applications such as the SPC/Fw water 
model~\cite{WuTepperVoth2006}, the 
interparticle potential contains, beyond the Coulomb term, bond-fluctuation, 
bending, and Lennard-Jones contributions. For large $N$, the $\bigObs{N \log 
N}$ 
Coulomb event rate dominates the local contributions, as was observed in 
simulations (see~\cite[Fig.~15]{Faulkner2018}). In ECMC, it can be expected 
that 
the Coulomb dipole factors have mixing and correlation times that are a factor 
of $ N^{1/3}/ \log N$ smaller than those for Coulomb atomic factors, although 
this has not been verified explicitly. \\

\subsubsection{Computation of Coulomb event rates}
\label{subsec:ECMCCoulomb}

The evaluation of the Coulomb potential and of its derivatives is a major 
challenge for molecular simulation. Molecular dynamics simultaneously requires 
the derivatives of the full Coulomb potential for all particles at each time 
step The dominant Ewald-sum-based solution to this problem consists in 
discretizing charges onto a fine spatial lattice and in solving Poisson's 
equation with fast Fourier transformation. The complexity of this algorithm is 
$\bigObs{N \log{N}}$ with a prefactor that depends on the numerical precision 
(see~\cite{FrenkelSmitBook2001} and~\cite[Sect. 1A]{Faulkner2018}) The 
difficulty in evaluating the difference of the Coulomb potential also 
frustrates 
energy-based local-move MCMC: Updating a single particle costs at least 
\bigObs{N^{1/2}} with known algorithms~\cite{BerthoumieuxMaggs2010} so that a 
sweep of local moves costs a prohibitive \bigObs{N^{3/2}}.

In ECMC with Coulomb interactions, a naive implementation of the factorized 
Metropolis algorithm would need to compute \bigObs{N} Coulomb factor event 
rates in order to decide on a possible veto of the consensus process. 
The thinning implemented in the cell-veto algorithm reduces this complexity to 
\bigObs{1}, because in a first part, the position-dependent Coulomb factor 
event rates are \quot{thickened} into time-independent and precomputed 
rates. Specific computations are required only to confirm a provisional 
veto. Approximate information on the location (and possibly the 
orientation) of molecules is sufficient in order to compute the 
(\quot{thickened}) bounding potentials and event 
rates, while the confirmation step accesses the factor event rates to machine 
precision. 

The separate-image and the merged-image formulations both lead to conditionally 
converging sums, that need to be regularized.  
In the  separated-image formulation of the Coulomb problem, 
counter-line-charges~\cite{KapferKrauth2016} and their generalizations as, for 
example, the counter-volume-charges~\cite[Fig. 5]{Faulkner2018}, regularize 
each individual image charge. Likewise, in the merged-image formulation, 
the sum over all images with associated compensating line charges is equivalent
to the standard tin-foil boundary condition~\cite{deLeeuw1980-1} for a single 
pair of Coulomb charges in a finite periodic box (see~\cite[Sect. 
III]{Faulkner2018}). \\

\subsubsection{ECMC liftings and lifting schemes for molecular simulation}
\label{subsec:ECMCLiftingSchemes}

As emphasized throughout this review, the factorized Metropolis algorithm 
attributes each MCMC rejection to a single factor and transforms it, in  
ECMC, into a lifting. For a pair factor, this lifting is usually
determined uniquely (the active and the target particles exchange roles). For a 
three-particle factor, the probability distribution for the new active particle 
is uniquely determined, whereas for larger factors, there is a choice among 
probability distributions, socalled \quot{lifting schemes}. For Coulomb dipole 
factors of two water molecules, each such factor contains six particles, and 
lifting schemes can  differ in their inside-flow 
(active particles on the same molecule for the in-state as for the out-state) 
and in their outside-flow. This is exemplified in the \quot{inside-first} and 
the \quot{outside-first} lifting schemes~\cite{Faulkner2018}.

For a factor in which the two molecules are separated by a distance $r$, the 
intra-molecular lifting rate scales as $1/r^3 $ for the inside-first lifting 
scheme whereas the lifting rate towards the other molecule scales as $1/r^4$.
Integrated over the whole simulation box of length $L$, the lifting rate inside 
the original molecule scales as $\log L$, whereas the outside lifting rate 
remains constant. For large $L$, the lifting remains inside the original 
molecule with probability $1$. For the outside-first lifting scheme, in 
contrast, both the intra-molecule and the inter-molecule lifting rates are 
$\sim 1/r^3$ as a function of the distance between molecules, and the total 
intro-molecule and inter-molecule lifting rates both  scale as $\sim \log L$ 
for large $L$~\cite[Table 1]{Faulkner2018}. In addition to the lifting schemes, 
the choice for the set $\LCAL$ and for the choice of selections of its 
elements, 
remains poorly understood~\cite{Qin2020fastsequential}.  \\

\subsection{Jellyfysh application for ECMC}
\label{subsec:Jellyfysh}

The open-source \quot{JeLLyFysh} application~\cite{Hoellmer2020} is the first 
general-purpose open-source implementation of ECMC. The configuration files of 
its Version 1.0 realize proof-of-concept of ECMC for interacting particle 
systems and for the handling of the Coulomb interactions.  They also showcase 
different factorizations and implement various liftings and lifting schemes as 
well as thinning strategies through bounding potentials and through the 
cell-veto algorithm. The application provides a platform for future method 
development, for benchmarking and for production code. The application's 
architecture (using the mediator design pattern) mirrors the mathematical 
structure of ECMC. Its event-driven nature is far removed from the time-driven 
setup of present-day molecular dynamics codes.  \\

\subsubsection{Mediators, activators, and event handlers}
\label{subsec:JellysyshMediatorsETC}

The \quot{JeLLyFysh} architecture is entirely based on the concept of 
events. A mediator~\cite{GammaDesignPatterns1994} serves as a central hub for 
all other elements of the code. After each event, this mediator accesses the 
active particles (in fact, the \quot{active global state}) and then fetches 
from 
another element, the activator, socalled event handlers that each treat one 
factor (depending on the active particle but also on the previous event). For 
each of these event handlers (that is, roughly, for each factor), the mediator 
then accesses the in-state for each factor, which allows the event handler to 
compute a candidate event time. All candidate event times are compared in an 
element of the program called a scheduler. For the shortest time in this 
scheduler, the corresponding event handler is then asked to compute an 
out-state. If confirmed, the out-state is 
committed to the state of the system. Output may also be generated (see 
\cite[Fig. 5]{Hoellmer2020}).

Besides those lifting events imposed by the global-balance condition, 
the JeLLyFysh application implements a number of pseudo-events. These 
pseudo-events correspond to sampling, for moving from one cell to another, for 
changing the direction of motion, \emph{etc.} and even 
for the start and the end of the program. It also takes into account the 
fundamental modularity of ECMC. For example, each event handler, for a 
factor $M$, only receives the factor in-state $c_M$ in 
\eq{equ:PotentialFactorized} in order to contribute a candidate event. It 
computes the  
out-state only if $M$ is the factor that causes the 
veto, that is, the true event. \\

\subsubsection{Configuration files, performance tests}
\label{subsec:JellyfyshTests}

The configuration files of  JeLLyFysh Version 1.0 construct runs for two 
charged 
point masses, for interacting charged dipoles, and for two interacting SPC/Fw 
water molecules, each with a large choice of factors, and high-precision 
comparison with reversible MCMC. Four different configuration files treat two
interacting SPC/Fw water molecules, with their harmonic oxygen--hydrogen bond 
interaction on one molecule, a three-body bending potential, a Lennard-Jones 
oxygen--oxygen potential as well as a Coulomb potential between atoms on 
different molecules~\cite[Fig. 12]{Faulkner2018}. The configuration files 
illustrate the possibilities offered to choose between atomic and molecular 
factors,  and to implement event-handlers that invert potentials to obtain 
direct event rates, to use bounding potentials, or even to invoke the cell-veto 
algorithm. Correlation functions are shown to agree in the $0.1 \% $ range. 

Performance tests of ECMC for large systems of SPC/Fw water molecules will 
be the object of the next version of the JeLLyFysh application. It features
substantially rewritten code, with many of the Python modules rewritten in a 
faster compiled language. Benchmarks will be provided against single-processor 
LAMMPS code. \\

\section{Prospects}
\label{sec:Prospects}

This article has reviewed the mathematical and algorithmic foundations of ECMC 
and analyzed a number of illustrative examples. It has also discussed first 
applications of the method to key models in statistical physics as well as an 
ongoing initiative to apply it in the field of molecular simulation. It is now 
time to resume the principal characteristics of ECMC, to formulate the main 
working assumptions behind its development, and to define its major challenges.

ECMC reinterpretes three key aspects of reversible Markov-chain Monte Carlo 
methods. First, the detailed-balance condition is replaced by global balance, 
in 
other words the condition of vanishing probability flows is replaced by a 
steady-state condition, which may allow a swifter exploration of sample space. 
Second, the energy-driven Metropolis algorithm is replaced by the factorized 
Metropolis algorithm, which is consensus-driven. Systematically, in ECMC, the 
consensus must be broken by a single factor, and this usually brings about its 
formulation as a continuous-time Markov chain. Once the vetoing factor is 
identified, it is sole responsible for the time evolution into and out of the 
next event.  Third, the trademark rejections of the Metropolis algorithm are 
reinterpreted as liftings, which enable persistent moves. ECMC has no 
rejections 
in the lifted sample space and, under certain symmetry conditions such as 
global 
translational invariance, allows one to separate certain lifting sectors from 
each other and to sample only one such sector.

The ongoing development of ECMC for molecular simulation and related fields 
builds on two assumptions. The first assumption is that Newtonian dynamics, 
while being physically realistic, is not the fastest possible relaxation 
algorithm towards the stationary distribution. The three-decade-long investment 
in molecular-dynamics codes will thus allow alternative approaches to be 
ultimately successful. The second assumption is that the elaborate \emph{a 
priori} choices that lead to global MCMC moves cannot be adapted to real-world 
applications in 
molecular simulation and related fields, because of the sheer complexity of the 
total potentials. This leaves one with local algorithms but, as argued 
throughout this review, they must be non-reversible in order to be fast. The 
family of ECMC algorithms provides a framework for non-reversible local MCMC. It 
opens up numerous opportunities, from the choice of factors and liftings to the 
new approach to the Coulomb problem that avoids the computation of the potential 
and its derivatives.

ECMC enlarges the possibilities of MCMC algorithm and is therefore guaranteed to 
be useful for specific applications in physics and other disciplines. 
One promising field of application is polymer 
science~\cite{KampmannKierfeldJCP2015,Mueller2020}. As a 
general method for molecular simulation, ECMC however faces two critical 
short-term challenges. The first concerns the benchmarks against molecular 
dynamics for standard real-world systems.  Such benchmarks, with 
single-processor codes for ECMC and molecular dynamics, will be the object of an 
upcoming version of the JeLLyFysh application. Their success will depend on the 
incorporation of appropriate lifting choices into this open-source code. The 
second challenge concerns the parallelization of ECMC. A road-map for 
multithreaded ECMC exists at present only for hard-sphere systems. The 
development of genuine parallel event-driven ECMC for generic potentials 
constitutes an open research subject. More generally, the mathematical 
understanding of the mixing and relaxation dynamics of reversible and 
non-reversible local MCMC remains rudimentary for next to all $N$-particle 
models in more than one dimension. A better mathematical grasp of such systems, 
beyond straightforward benchmarks, appears as a longer-term research goal as 
well as a prerequisite for the development of faster algorithms that are 
required for applications in physics and throughout the  sciences. \\

\section*{Funding}
The author acknowledges support from the Alexander von Humboldt Foundation.  \\

\section*{Acknowledgments}
The author is indebted to S. C. Kapfer for discussions on 
\sect{sec:SingleParticles}  and to  S. Todo for discussions on 
\sect{subsec:StochasticPotentialSwitching}. \\

\bibliographystyle{frontiersinSCNS_ENG_HUMS} 

\bibliography{/home/krauth/PDF_papers/WernerKrauthGeneral.bib,test.bib}

\end{document}

%% file: latexcommands_2021a.tex
\newcommand{\SET}[1]{\{#1\}}

\newcommand{\eq}[1]{Eq.~\eqref{#1}}
\newcommand{\eqtwo}[2]{Eqs~\eqref{#1} and~\eqref{#2}}

\newcommand{\quot}[1]{``#1''}

\newcommand{\sect}[1]{Section~\ref{#1}} 
\newcommand{\subsect}[1]{Subsection~\ref{#1}}

\newcommand{\secttwo}[2]{Sections~\ref{#1} and~\ref{#2}}

\newcommand{\ACAL}{\mathcal{A}}  
\newcommand{\DCAL}{\mathcal{D}}  
\newcommand{\FCAL}{\mathcal{F}}  
\newcommand{\LCAL}{\mathcal{L}}  
\newcommand{\MCAL}{\mathcal{M}}  
\newcommand{\OCAL}{\mathcal{O}}  
\newcommand{\PCAL}{\mathcal{P}}  
\newcommand{\RCAL}{\mathcal{R}}  
\newcommand{\Sbar}{\overline{S}}  
\newcommand{\Ubar}{\overline{U}}  
\newcommand{\maxZeroa}[1]{\gla #1 \gra^+}   
\newcommand{\maxZeroc}[1]{\glc #1 \grc^+}   

\newcommand{\expb}[1]{\exp \glb #1 \grb} 
\newcommand{\expc}[1]{\exp \glc #1 \grc} 
\newcommand{\expd}[1]{\exp \gld #1 \grd} 
\newcommand{\ranb}[2][]{\ran_{#1} \! \glb #2 \grb}  

\newcommand{\cosb}[2][]{\cos^{#1} \glb #2 \grb}  

\newcommand{\loga}[2][]{\log^{#1}\! \gla #2 \gra}  
\newcommand{\logb}[2][]{\log^{#1} \glb #2 \grb}  

\newcommand{\probb}[1]{\mathbb{P} \glb #1 \grb}

\newcommand{\gla}{\,}  
\newcommand{\gra}{}  
\newcommand{\glb}{\left(}  
\newcommand{\grb}{\right)}  
\newcommand{\glc}{\left[}  
\newcommand{\grc}{\right]}  
\newcommand{\gld}{\left\{}  
\newcommand{\grd}{\right\}}  

\newcommand{\const}{\text{const}}

\newcommand{\TO}{,\ldots,}
\newcommand{\VEC}[1]{\mathbf{#1}}
\newcommand{\rvec}{\VEC{r}}

\newcommand{\xvec}{\VEC{x}}

\newcommand{\Utilde}{\tilde{U}}

\newcommand{\xtilde}{\tilde{x}}

\newcommand{\deltavec}{\boldsymbol{\delta}}

\newcommand{\chat}{\hat{c}}

\newcommand{\Phat}{\hat{P}}

\newcommand{\pihat}{\hat{\pi}}

\newcommand{\Omegahat}{\hat{\Omega}}

\newcommand{\neigh}[2]{\langle #1 , #2 \rangle}

\newcommand{\mean}[1]{\left\langle #1 \right\rangle}
\newcommand{\half}{\frac{1}{2}}

\newcommand\bigObs[1]{\ensuremath{\OCAL  (#1)}}

\newcommand\diff[1]{\mathrm{d}#1}

\DeclareMathOperator{\ran}{ran}